\documentclass[fleqn,usenatbib]{mnras}

\usepackage{newtxtext,newtxmath}
\usepackage[T1]{fontenc}
\usepackage{soul}

\DeclareRobustCommand{\VAN}[3]{#2}
\let\VANthebibliography\thebibliography
\def\thebibliography{\DeclareRobustCommand{\VAN}[3]{##3}\VANthebibliography}

\usepackage{graphicx}	% Including figure files
\usepackage{amsmath}	% Advanced maths commands
\usepackage{caption}
\usepackage{float}
\usepackage{color}
\usepackage[dvipsnames]{xcolor}
\usepackage{subcaption}
\usepackage{placeins}
\usepackage{listings}
\usepackage[framemethod=default]{mdframed}
\mdfdefinestyle{theoremstyle}{% 
linecolor=black,linewidth=2pt,% 
frametitlerule=true,% 
frametitlebackgroundcolor=gray!20, 
innertopmargin=\topskip,}
\newcommand{\tC}{{$\rm \theta^1$~Ori~C}}
\newcommand{\tX}{{$\rm \theta^2$~Ori~A}}
\newcommand{\kms}{km~s{$^{-1}$}}

\newcommand{\heI}{He I}
\newcommand{\Hb}{H$\beta$}
\newcommand{\Ha}{H$\alpha$}
\newcommand{\Lyb}{Ly$\beta$}
{\Large {\Large }}

\newcommand{\oi}{[\ion{O}{I}]}
\newcommand{\oiii}{[\ion{O}{III}]}

\newcommand{\nii}{[\ion{N}{II}]}
\newcommand{\sii}{[\ion{S}{II}]}

\newcommand{\On}{O$\rm ^{o}$}

\newcommand{\Ci}{[\ion{C}{I}]}

\title[Outflows and discs with MUSE]{The VLT MUSE NFM view of outflows and externally photoevaporating discs near the Orion Bar\thanks{Based on observations collected at the European Southern Observatory under ESO programme 110.259E.001.}}
\author[Haworth et al.]{Thomas J. Haworth$^1$\thanks{t.haworth@qmul.ac.uk}, Megan Reiter$^2$, C. Robert O'Dell$^3$, Peter Zeidler$^4$, Olivier Berne$^5$, Carlo F. Manara$^6$, 
\newauthor 
Giulia Ballabio$^7$, Jinyoung S. Kim$^8$, John Bally$^9$, Javier R. Goicoechea$^{10}$, Mari-Liis Aru$^6$, Aashish Gupta$^6$ \newauthor and Anna Miotello$^6$%, Aashish Gupta$^X$ and    Anna Miotello$^Y$
\\
$^{1}$Astronomy Unit, School of Physics and Astronomy, Queen Mary University of London, London E1 4NS, UK\\
$^{2}$Department of Physics and Astronomy, Rice University, 6100 Main St - MS 108, Houston, TX 77005, USA\\
$^3$Department of Physics and Astronomy, Vanderbilt University, Nashville, TN 37235-1807, USA \\
$^{4}$AURA for the European Space Agency (ESA), ESA Office, Space Telescope Science Institute, 3700 San Martin Drive, Baltimore, MD 21218, USA \\
$^{5}$ Institut de Recherche en Astrophysique et Plan\'etologie, Universit\'e de Toulouse, CNRS, CNES, UPS, 9 Av. du colonel Roche, 31028 Toulouse Cedex 04, France \\
$^6$ European Southern Observatory, Karl-Schwarzschild-Strasse 2, 85748 Garching bei München, Germany \\
$^7$ Astrophysics Group, Imperial College London, Blackett Laboratory, Prince Consort Road, London SW7 2AZ, UK \\
$^8$ Steward Observatory, University of Arizona, 933 N. Cherry Ave, Tucson, AZ 85721-0065, USA \\
$^9$ Center for Astrophysics and Space Astronomy, Astrophysical and Planetary Sciences Department, University of Colorado, UCB 389
Boulder, Colorado 80309, USA \\
$^{10}$ Instituto de F\'{\i}sica Fundamental (CSIC). Calle Serrano 121-123, 28006, Madrid, Spain.\\
}

\date{Accepted XXX. Received YYY; in original form ZZZ}

\pubyear{2023}

\begin{document}
\label{firstpage}
\pagerange{\pageref{firstpage}--\pageref{lastpage}}
\maketitle

\begin{abstract}
We present VLT/MUSE Narrow Field Mode (NFM) observations of a pair of disc-bearing young stellar objects towards the Orion Bar: 203-504 and 203-506.  Both of these discs are subject to external photoevaporation, where winds are launched from their outer regions due to environmental irradiation. Intriguingly, despite having projected separation from one another of only 1.65\arcsec (660\,au at 400\,pc), 203-504 has a classic teardrop shaped ``proplyd'' morphology pointing towards $\theta^2$Ori A (indicating irradiation by the EUV of that star, rather than \tC) but 203-506 has no ionisation front, indicating it is not irradiated by stellar EUV at all. However, 203-506 does show \Ci\ 8727 \AA\ and \oi\ 6300 \AA\ in emission, indicating irradiation by stellar FUV.
This explicitly demonstrates the importance of FUV irradiation in driving mass loss from discs. 
We conclude that shielding  of 203-506 from EUV is most likely due to its position on the observers side of an ionized layer lying in the foreground of the Huygens Region.   
We demonstrate that the outflow HH 519, previously thought to be emanating from 203-504 is actually an irradiated cloud edge and identify a new compact outflow from that object approximately along our line of sight with a velocity $\sim130$\,km\,s$^{-1}$. 
\end{abstract}


%%%%%%%%%%%%%%%%%%%%%%%%%%%%%%%%%%%%%%%%%%%%%%%%%%

%%%%%%%%%%%%%%%%% BODY OF PAPER %%%%%%%%%%%%%%%%%%

\section{Introduction}

With ever increasing numbers of exoplanets being discovered \citep{2021exbi.book....2G}, there is a need to understand how the planet formation process contributes to exoplanet diversity. There is now overwhelming evidence that planets form from circumstellar ``planet-forming'' discs, to the extent that we can even directly observe young planets within them in some cases \citep{2018ApJ...860L..13P, 2018ApJ...860L..12T, 2018A&A...617A..44K, 2019NatAs...3..749H, 2022arXiv220309528P}. Our most detailed studies of these have typically been targeting large radius discs in the most nearby star forming regions, which are the easiest to detect and resolve features in. Those have provided great insight into the gas and dust composition and conditions, revealing substructures that may possibly be associated with the planet formation process, or planets themselves \citep[e.g.][]{2015ApJ...808L...3A, 2015MNRAS.453L..73D, 2018ApJ...869L..41A,2018ApJ...869L..47Z, 2020ApJ...890L...9P, 2021ApJS..257....1O}. While those observations of nearby bright discs are extremely important, recently there is a shift towards trying to study more typical planet-forming discs. For nearby regions, this entails studying in detail those of more typical radii\citep[$R\leq$50\,au,][Miley et al. submitted, and the DECO ALMA large program: 2022.1.00875.L, PI: Cleeves]{2023arXiv230503862Z,2021A&A...651A..48M}. However, for a truly general understanding of planet-forming disc evolution and planet formation we need to look to more distant, more massive, stellar clusters. 

The Sun is currently located in the middle of a supernova driven bubble that probably induced our nearest ($\sim140$\,pc distant) instances of star formation \citep[for example, in Taurus and Lupus][]{2022Natur.601..334Z}. These are all low mass star forming regions, containing of order hundreds of stars, all of which are $<10$\,M$_\odot$. The observed cluster mass function of $dN/dM \propto M^{-2}$ implies that most stars form in much larger massive clusters \citep[e.g.][]{2003ARA&A..41...57L, 2009Ap&SS.324..315C, 2010ApJ...710L.142F}. Massive star clusters form from the clumpy/hierarchical collapse (or merger) of turbulent giant molecular clouds \citep{2003ARA&A..41...57L, 2010ARA&A..48..431P, 2021PASJ...73S...1F}. Once massive stars form their radiative (photoionisaton/radiation pressure) and mechanical (winds/supernovae) feedback heats and disperses the cloud \citep{2009ApJ...694L..26G, 2010ApJ...723..971G, 2011MNRAS.414..321D, 2012MNRAS.424..377D, 2014MNRAS.442..694D, 2012MNRAS.427..625W, 2021MNRAS.506.2199G, 2022MNRAS.512..216G, 2022MNRAS.515.4929G}. The high stellar number density, dispersal of the molecular cloud and high UV radiation field strengths all mean that the star forming environment in which the planet-forming discs reside is very different to lower mass star forming regions.

Massive stellar clusters have high stellar number densities, so the probability of gravitational encounters between star-disc systems is more likely \citep{2005ApJ...629..526P, 2013A&A...549A..82P, 2018ApJ...863...45P, 2019MNRAS.483.4114C, 2020MNRAS.491..504C, 2023EPJP..138...11C}. Additionally, OB stars  dominate the production of UV photons by many orders of magnitude (for example a 40\,M$_\odot$ star emits around ten trillion times more ionising photons than a Solar type star). One impact of that is molecular cloud dispersal, which may influence the timescale that discs can be fed by accretion streamers \citep[e.g.][]{2012A&A...547A..84T, 2016ARA&A..54..271K, 2019ApJ...880...69Y, 2020NatAs...4.1158P,  2021ApJ...908L..25G, 2022A&A...658A.104G, 2022A&A...667A..12V, 2022ApJ...930..171H,   2023A&A...670L...8G}. The much more established impact of the UV from massive stars is that it leads to ``external'' photoevaporative winds being launched from the outer parts of irradiated discs \citep[for recent reviews see][]{2022EPJP..137.1132W, 2022EPJP..137.1071R}. External photoevaporative winds are spatially resolvable in relatively nearby sources ($D\sim400$\,pc), with teardrop shape ionisation front morphologies now referred to as ``proplyds'' that point towards the UV source responsible for irradiating them \citep[e.g.][]{1993ApJ...410..696O, 1994ApJ...436..194O,1999AJ....118.2350H,  2000AJ....119.2919B, 2008AJ....136.2136R, 2012ApJ...757...78M, 2016ApJ...833L..16F, 2016ApJ...826L..15K, 2021MNRAS.501.3502H, 2022EPJP..137.1132W}. Proplyd mass loss rates are empirically inferred to be sufficiently high that it would result in rapid depletion and truncation \citep{1999AJ....118.2350H,2019MNRAS.490.5478W, 2022EPJP..137.1132W} and  models support these high mass loss rates \citep{2000ApJ...539..258R, 2004ApJ...611..360A, 2016MNRAS.457.3593F, 2018MNRAS.481..452H, 2019MNRAS.485.3895H}. 
Recent models predict that external photoevaporation could impact  planet formation anywhere in the disc, by reducing the reservoir of solids that can radially drift and resupply the inner disc with material \citep{2023MNRAS.522.1939Q}, as well as affect the migration and growth of giant planets. The discs of  proplyds have also been subject to detailed study to understand the impact of the external winds upon them, finding that they are typically more compact, less massive and even chemically distinct from discs in low mass star forming regions \citep[e.g.][]{2014ApJ...784...82M, 2018ApJ...860...77E, 2020ApJ...894...74B, 2023ApJ...947....7B}.

Because the impact of external photoevaporation on discs is expected to be so significant, it makes the connection to the wider star formation process particularly important. Stars form embedded and hence shielded from external photoevaporation until they dynamically emerge from the molecular cloud, or the cloud itself is dispersed by feedback. This has been shown to be important for understanding disc properties (masses, radii, lifetimes) in clusters \citep{2022MNRAS.512.3788Q, 2023MNRAS.520.5331W, 2023MNRAS.520.6159C}, and the shielding timescale is predicted to have a significant impact on the resulting planetary architecture \citep{2023MNRAS.522.1939Q}. 

In this paper we study two externally irradiated discs coincident with the Orion Bar. One of these is a classic example of an externally photoevaporating disc, with an extreme ultraviolet (EUV) excited teardrop shaped ionisation front. The other is somehow shielded from EUV irradiation, has a wind driven purely by far-ultraviolet (FUV). This pair of objects is valuable for understanding the role of external photoevaporation outside of the area of most intense feedback in the Orion Nebular Cluster, and the role that intervening material plays in changing the UV field incident upon discs.

\section{The Orion Nebula and the Orion Bar}
\begin{figure*}
    \centering
    \includegraphics[width=1.6\columnwidth]{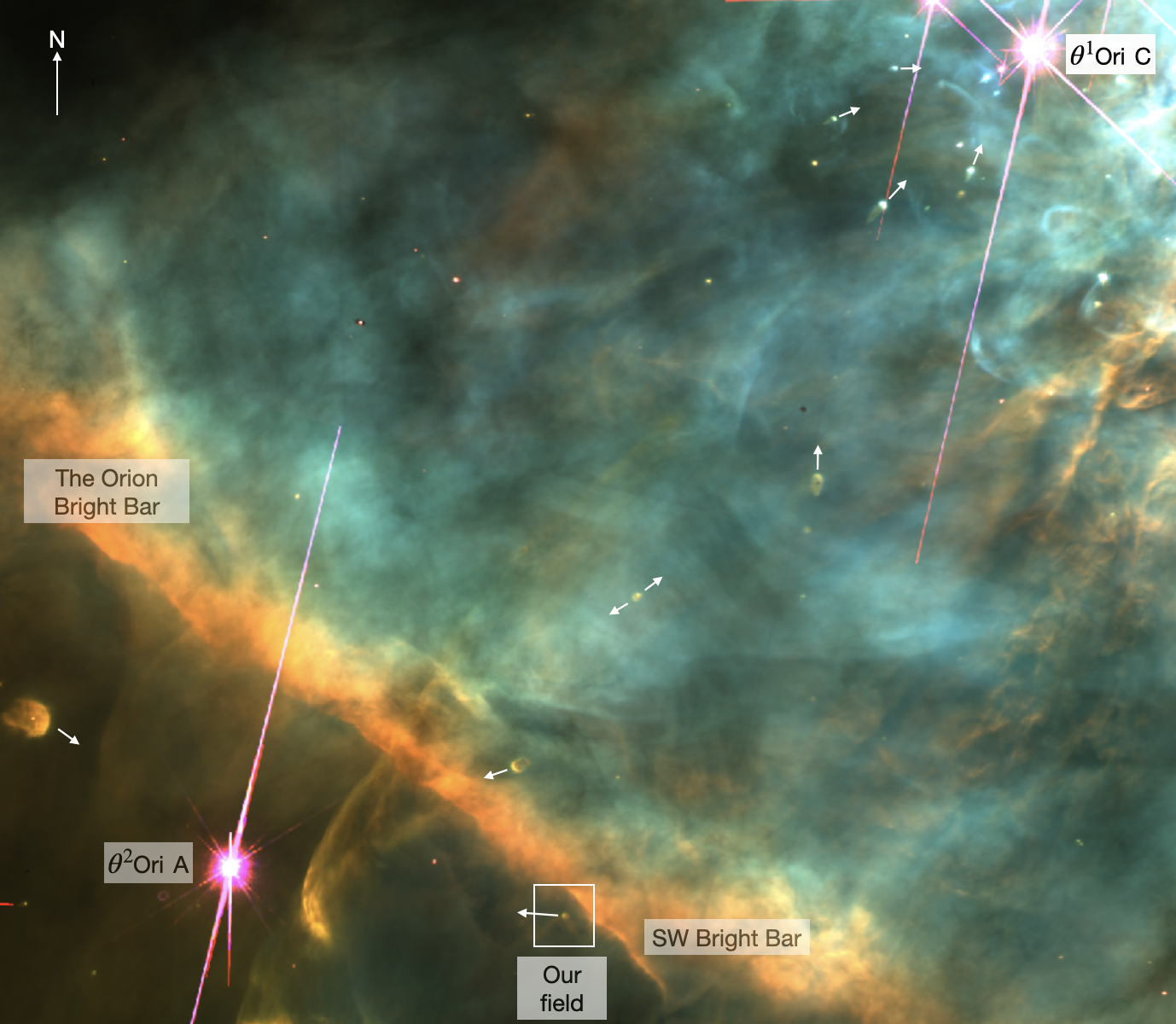} 
    \caption{A 3-colour HST [N\,\textsc{ii}] (red) H\,$\alpha$ (green) and [O\,\textsc{iii}] 5007 \AA\ image image of the Trapezium (upper right) and Orion Bar (also known as the Orion Bright Bar). The FOV that we study with VLT MUSE is the boxed region in the south western Orion Bar (which \protect\cite{2017MNRAS.464.4835O} call the SW Bright Bar). The orientation of ionisation fronts associated with the most prominent proplyds is denoted by the white arrows.  The main UV source in the region is $\theta^1$\,Ori C in the Trapezium, but $\theta^2$\,Ori A is also clearly irradiating discs and has been shown to dominate UV irradiation of parts of the cloud south east of the Orion Bar \protect\cite{2017ApJ...837..151O}. Note that North is up in this image.  }
    \label{fig:ONC}
\end{figure*}

 At a distance of $\sim400\,$pc \citep{2017ApJ...834..142K, 2018AJ....156...84K}, the Orion Nebula Cluster is the nearest and best studied massive star forming region.  
The primary source of UV photons in the region is from the Trapezium stars, and of those most notably the O6V star $\theta^1$ Ori C \citep{2017ApJ...837..151O}. Figure \ref{fig:ONC} shows a Hubble Space Telescope 3-colour ([N\,\textsc{ii}], H\,$\alpha$, [O\,\textsc{iii}]) image of the overall region, with the main features labeled, including the location of the main field of view (FOV) that we consider in this paper.  This HST image is made from the mosaic of emission-line filters (red-[N II] 6583 \AA, green-H$\,\alpha$, blue-[O III] 5007 \AA) first presented in \cite{1996AJ....111..846O}, but reprocessed and improved by Jeremy R. Walsh and Richard Hooks of the European Southern Observatory. 

The overall ionisation structure around $\theta^1$ Ori C  is as follows. The innermost photoionised region contains ionised hydrogen (ionisation energy 13.6\,eV) and singly ionised helium (ionisation energy 24.59\,eV). There is no doubly ionised helium zone because $\theta^1$ Ori C does not emit enough sufficiently energetic ($>54.4$\,eV) photons. This singly ionised helium zone is most easily traced by the [O\,\textsc{iii}] 5007 \AA\ emission line from doubly ionised oxygen. Moving outwards through the photoionised gas, all of the photons capable of ionising helium are absorbed, resulting in a zone of neutral helium but ionised hydrogen that is most easily traced by [N\,\textsc{ii}] 6583 \AA\ emission. At the distance where all of the hydrogen ionising photons are used in maintaining photoionisation balance in the H\,\textsc{ii} region is a relatively sharp ionisation front, beyond which is a photodissociation region (PDR).

The brightest area of the Orion Nebula is historically called the Huygens Region. 
Quantitative analysis of emission lines there has led to the interpretation of the Huygens Region being a concave surface with irregularities\citep{1995ApJ...438..784W,2005ApJ...627..813H,2009AJ....137..367O,2017ApJ...837..151O}. 

 In this picture, the Trapezium stars are near the more distant side of the concave surface, which slopes closer to the observer towards a bright, linear southeast feature. The {main ionisation front (MIF) associated with this feature is where the hydrogen ionisation front is along the line of sight. In the foreground is a ``nearer ionised layer''  \citep[NIL][]{2020ApJ...891...46O}. The density and temperature structure of the Huygens Region has been well studied \citep[e.g.][]{2015A&A...582A.114W, 2016MNRAS.455.4057M, 2017MNRAS.464.4835O}. 

The Huygens region of the Orion Nebula contains multiple elongated bar structures \citep{2017MNRAS.464.4835O} but the most famous, around  $112''$ to the south east of the Trapezium is the NE-SW tending Orion  Bar (alternatively referred to as ``the Bright Bar" in many visual studies, see Figure \ref{fig:ONC}). In this report we will subsequently call it the ``Bar''. The Bar is a feature caused by the MIF of the central Orion Nebula being tilted so much that it is viewed edge-on. This allows a stacked view of the structure in the underlying PDR. The Bar has been the subject of a number of 1D modelling efforts \citep[e.g.][]{1989ApJ...344..791B, 1994ApJ...422..136T, 1995A&A...303..541J, 2000A&A...364..301W, 2015A&A...575A..82C}, as well as models and observations exploring the importance of clumpiness and 3D structure in PDRs \citep{2017A&A...598A...2A}. \cite{2016Natur.537..207G} find that considering the Orion bar as a static slab may be insufficient, since it is being compressed and ablated by the high pressure photoionised gas and/or photoevaporative compression by the rocket effect. \cite{2023A&A...673A.149H} also demonstrate the complicated sub-structured nature of the Orion Bar.

The region that we study in this paper contains two externally photoevaporating young stellar objects (YSOs), 203-506 and 203-504, at the boundary of the Bar. Their coordinates are given in Table \ref{tab:coords}. 
Both objects have an associated outflow: HH~519 and HH~520, respectively. 
HST has already demonstrated that 203-504 is a proplyd that is directed towards $\theta^2$ Ori A \citep{2000AJ....119.2919B, 2017ApJ...837..151O}.
203-506 was included in the Herschel study of \citet{2017A&A...604A..69C}, \cite{2023A&A...673A.149H} recently detected vibrationally excited  (v=1-0 S(1) at 2.12\,$\mu$m) H$_2$ emission,
and Berne et al. (submitted) have most recently used  JWST \citep{2022PASP..134e4301B} and ALMA to study 203-506.  

Our objective has been to use the European Southern Observatory (ESO) integral field spectrograph MUSE on the Very Large Telescope (VLT), which in the Narrow Field Mode (NFM) has spatial resolution comparable to HST and spectral coverage over the entire optical range to understand more about these apparently very different externally photoevaporating discs, which are  seen in projection close to the boundary of the Huygens Region, just beyond the  peak brightness of the Bar. The multiplicity of emission-lines in the visual region were expected to give a better depiction of the environment of both 203-504 and 203-506 and a refinement of the physical conditions within 203-506. 

\begin{table}
    \centering
    \begin{tabular}{cccc}
    \hline    
    System  & RA  & DEC &  \\
    \hline     
    203-504 (HH 519)     & 05 35 20.26 &-05 25 04.05 & \\    
    203-506 (HH 520)     & 	05 35 20.32 & -05 25 05.55 & \\  
    \hline
    \end{tabular}
    \caption{The coordinates of the YSOs 203-504 and 203-506 \citep{2008AJ....136.2136R}.}
    \label{tab:coords}
\end{table}

\subsection{The distance to $\theta^2$ Ori A}
\label{ref:distThet2OriA}
Although not the dominant UV source in the Orion Nebular Cluster (ONC), the O9.5V star  $\theta^2$ Ori A  plays an important role in this paper. It is located to the south east of the Bar (it is labeled in Figure \ref{fig:ONC}).  The details of our interpretation are sensitive to the distance of  $\theta^2$ Ori A relative to the rest of the cluster. The distance to the ONC is generally inferred to be around 390-400\,pc \citep{2007MNRAS.376.1109J, 2017ApJ...834..142K,2018AJ....156...84K, 2018A&A...619A.106G}. The Gaia DR3 and DR2 distances to $\theta^2$ Ori A are  336 and 450\,pc respectively \citep{2016A&A...595A...1G, 2018A&A...616A...1G, 2022arXiv220800211G} so are wildly inconsistent with one another, suggesting that the current Gaia estimates are affected by some factor such as its multiplicity and/or larger uncertainties in the parallax of bright sources \citep{2006ApJ...653..636S} or nebulosity of the region. Given its importance and our inability to use a Gaia distance in this case we review the other body of evidence on its location, which supports $\theta^2$ Ori A being at a similar distance as the main cluster. 

The strongest argument lies in the two-pronged study of \citet{2017ApJ...837..151O} which addresses the sources of ionization of the Orion Nebula and its MIF. There they established that the bright ionized rims of the proplyds near \tC\ were oriented towards that star. However, when proceeding to the SE, the proplyd bright rims transition to orientation towards both \tC\ and \tX, until becoming exclusively oriented towards \tX\ (see Figure \ref{fig:ONC}, where white arrows denote the proplyd orientations).  They also examined the line ratios of the brightest lines in both the visual and infrared windows, using the  \textsc{cloudy} code \citep{1998PASP..110..761F, 2013RMxAA..49..137F, 2017RMxAA..53..385F} and the known radiation properties of the stars. They found the MIF in the Huygens Region to be photoionized by \tC, but to the southeast of the Bar, \tX\ was the source of ionization.

 The radial velocity of \tX\ is compatible with its membership in the ONC. Although the SIMBAD data-base gives this as 19 \kms\ citing an early low resolution study \citep{1920PDO.....5....1H}, examination of that paper shows that although the star was recognized as a spectroscopic binary, the results were confused by the presence of emission lines in the spectra. A definitive value for the system velocity was determined from the orbital solution of \cite{1991ApJ...367..155A} to be 24.8$\pm$2.1 \kms. This is essentially the same as the most frequent velocity of the ONC stars of 25$\pm$2 \kms\ in the study of \cite{2005AJ....129..363S}. This agreement would not be expected if \tX\ was not a cluster member. Note that all velocities in this paper are heliocentric, for LSR subtract 18.2 \kms, which is arrived at under the usual assumption that the Sun is moving at 20 km/s towards 18:00:00 +30:00:00 (epoch 1900).

Overall the above arguments all point towards $\theta^2$ Ori A not being significantly in the foreground, rather, being close enough to the nebula to play a role in its photoionization processes.

\section{Observations and data reduction}
We were awarded 1 hour of Directors Discretionary Time (DDT) on VLT MUSE (PI: Haworth, run ID:110.259E.001). MUSE (Multi Unit Spectroscopic Explorer) is the visual wavelength integral-field spectrograph on the VLT mounted on UT4 \citep{2010SPIE.7735E..08B}. Our observations were a single MUSE narrow field mode (NFM) pointing with adaptive optics (AO). The observations were taken in service mode on the 20th November 2022 with 4 integrations of 11 minutes, giving 44 minutes on source. 4 raw datasets were provided in the observing block at 90 degree orientations to one another. These were reduced separately and then combined to make a single datacube. We reduced the data using the MUSEpack \citep{2019AJ....158..201Z, 2019zndo...3433996Z} python wrapper for the standard esorex (v2.8.7) framework \citep{2013A&A...559A..96F}. This utilised the reduced master calibration files provided by ESO. 

The FOV of the data is approximately $7.42\arcsec$ per side with an effective angular resolution of $0.0921\arcsec$, corresponding to spatial scales of 2968\,au and 36.8\,au respectively. The wavelength range spans 4750-9350\,\AA\, and the spectral resolving power over that range varies from $R=2000-4000$, corresponding to velocity resolution of $\sim75-150$\,km\,s$^{-1}$. We hence do not generally expect to resolve the kinematics of our targets, except possibly in the case of high velocity outflows aligned along the line of sight. 

The sky subtraction in the reduction of MUSE data uses either a dedicated sky observation, or the darkest part of the image. However, for bright and extended targets such as the Orion star forming region it is not possible to obtain an appropriate sky measurement because the telescope has to move by many degrees to find an appropriately dark field. In regions with high background, such as an H\,\textsc{\,\textsc{ii}} region, using the darkest part of an image may lead to over subtraction of key (nebular) lines of interest. To resolve this, we use the modified sky subtraction approach introduced by \cite{2019AJ....158..201Z}, which is included in MUSEpack. 
This assumes that the H\,\textsc{\,\textsc{ii}} region lines and continuum dominate over the sky, and uses OH and O$_2$ to determine and subtract the sky contribution. Even though this method may lead to a small overestimation of the source line fluxes (originated in line emission of Earth's atmosphere), it has been shown that overall this method performs better in terms of line profiles, flux ratios, and absolute wavelength (velocity) measurements \citep{2019AJ....158..201Z}.

To study lines we extract data $\pm3$\AA\ from the nominal line centre. The continuum subtraction uses nearby line free channels, typically 10 \AA\ from the line centre of interest so long as it is clear of other line emission. 

To correct for reddening we follow \cite{2010AJ....140..985O, 2015A&A...582A.114W, 2019MNRAS.490.2056R} and use the Balmer decrement to estimate the line of sight extinction. That is, we assume an intrinsic H$\,\alpha$/H\,$\beta$ ratio of 2.89 \citep{1995MNRAS.272...41S, 2006agna.book.....O} and use the \textsc{pyneb} \citep{2015A&A...573A..42L} reddening correction routine with a \cite{1989ApJ...345..245C} and  \cite{2007ApJ...655..299B} extinction law and assuming an optical total-to-selective extinction ratio $R_V=5.5$ \citep{1989ApJ...345..245C, 1991ApJ...374..580B, 1992ApJ...389..305O, 1994A&A...284..919G, 2007ApJ...655..299B, 2015A&A...582A.114W, 2021ApJ...908...49F}. That extinction law and  higher than standard value of $R_V$ for the diffuse ISM was found to be more appropriate for the ONC in a VLT MUSE study of hundreds of stars in the ONC by \citep{2021ApJ...908...49F}. A value of $R_V=5.5$ was also adopted in the MUSE study of the ONC by \cite{2015A&A...582A.114W}.

\section{Results}

\begin{figure}
    \centering
     \includegraphics[width=\columnwidth]{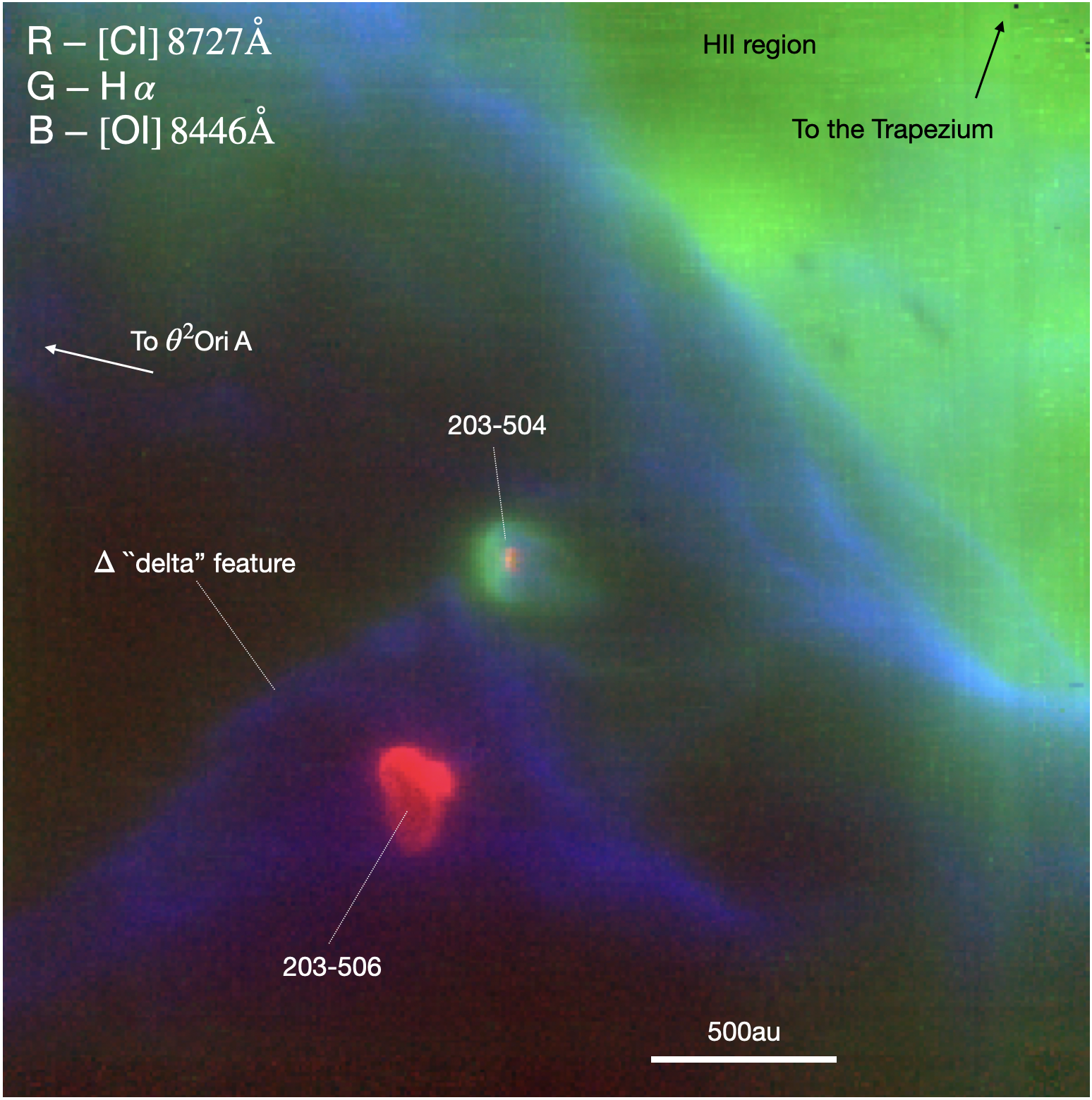}
    \caption{A 3 colour composite image from our dataset consisting of [C\,\textsc{i}] 8727 \AA\ (red), H$\,\alpha$ (green) and [O\,\textsc{i}] 8446 \AA\ (blue). The two disc-bearing YSO's are 203-504 and 203-506. To the upper left in green is the H\,\textsc{ii} region that is photoionised by the Trapezium stars (predominantly the O6.5V star $\theta^1$ Ori C). The O9.5V star $\theta^2$ Ori A is towards the left of the FOV. Blue traces Lyman $\beta$ irradiated gas, which is predominantly at the interface with the H\,\textsc{ii} region, but also traces an irradiated cloud near the line of sight location of the two YSOs, which based on its shape we call the Delta ($\Delta$) Feature.  }
    \label{fig:MUSEoverview}
\end{figure}

\subsection{Overview}
\label{sec:overview}
We begin with an overview of our dataset and its key components. Figures \ref{fig:MUSEoverview} and  \ref{fig:MUSEoverview2}  show colour composites, and Figure \ref{fig:gallery} shows a gallery of individual continuum subtracted emission lines. 

Figure \ref{fig:MUSEoverview} combines [C\,\textsc{i}] 8727 \AA\ (red), H\,$\alpha$ (green) and [O\,\textsc{i}] 8446 \AA\ (blue). We will later argue in Appendix \ref{sec:appCi} that the [C\,\textsc{i}] 8727 \AA\ line traces recombinations in FUV irradiated gas, and in our field is predominantly tracing the circumstellar material of 203-506. H\,$\alpha$  is a well established tracer of photoionised gas. The [O\,\textsc{i}] 8446 \AA\  traces Lyman $\beta$ irradiated gas because H\,\textsc{i} Lyman $\beta$ radiation ($\lambda=1025.72$\,\AA) has a very similar wavelength to the 2$\rm p^{3}$ 3d $\rm ^{3}D^{o}_{3}$ ground state of atomic oxygen ($\lambda=1025.76$\,\AA). The [O\,\textsc{i}] 8446 \AA\ line is a forbidden line emitted as part of the decay following resonant absorption of the Lyman $\beta$ as described by \cite{1928ApJ....67....1B} and also detailed in Figure 4.7 of \cite{2006agna.book.....O}. Usually the 8446 \AA\ line is weak in Huygens Region spectra, but it is often strong in our spectra. We have adopted the \cite{1928ApJ....67....1B} process described above as the source of the emission, with its requirement that \On\ is being irradiated by hydrogen \Lyb\ radiation \citep[see also][]{2006agna.book.....O}. 

The most striking features in Figure \ref{fig:MUSEoverview} are the two externally irradiated protoplanetary discs, 203-504 and 203-506. 203-504 exhibits the classic teardrop shaped  H\,$\alpha$ morphology of proplyds, pointing towards $\theta^2$\,Ori A. 203-506 is distinct, showing no clear ionisation front. We will discuss the very different external photoevaporation of these discs further in section \ref{sec:proplyds}.  

Figure \ref{fig:MUSEoverview2} again shows the [C\,\textsc{i}] 8727 \AA\ (red) and [O\,\textsc{i}] 8446 \AA\ (blue), but the green in this case is the [Fe~\textsc{ii}] 8617 \AA\ line. [Fe~\textsc{ii}] emission is well-established as a tracer of shock-excited gas in protostellar jets \citep[e.g.][]{2003A&A...410..155P,2004ApJ...614L..69H,2011A&A...532A..59A,2013ApJ...778...71G} and has also been observed in a number of externally-irradiated jets in H~{\sc ii} regions \citep[e.g.][]{2013MNRAS.433.2226R, 2015MNRAS.450..564R, 2016MNRAS.463.4344R}. In particular, \citet{2023A&A...673A.166K} used several [Fe~\textsc{ii}] lines observed with MUSE NFM to trace the properties of the jet from the 244-440 proplyd, also irradiated from $\theta^2$\,Ori A.
In our dataset [Fe~{\sc ii}] traces outflows from both 203-506 (HH 520) and 203-504 (HH 519), which we discuss in more detail in \ref{sec:jetsoutflows}.

To the upper right in Figure~\ref{fig:MUSEoverview} is the Huygens Region, photoionised by the Trapezium stars and bright in H~$\alpha$.The sharp boundary indicates where the MIF is nearly along the line-of-sight. The MIF then flattens into the lower surface-brightness region we call the Bar Shoulder and continues beyond to the southeast with progressively lower surface-brightness.  The low ionization boundary of the MIF is visible in Lyman $\beta$ irradiated [O\,{\sc i}] 8446 \AA. To the south-east, beyond the Bar ionisation front is the Delta Feature, that is enhanced in \oi\ 8446 \AA.  203-506 is along the line of sight of the Delta Feature, appearing in silhouette in many emission-lines (see Figure \ref{fig:gallery}) and 203-504 is located slightly to the northwest of the northern vertex of the Delta Feature. We discuss the possible role that this feature plays in the context of the externally photoevaporating discs in section \ref{sec:3dstructure}. Table \ref{tab:lines} provides a non-exhaustive summary of bright lines in our dataset and the features that they trace. 

Finally, within the H\,\textsc{ii} region in the upper right of Figure \ref{fig:MUSEoverview} are three small dark clumps, which are instrumental artefacts. We discuss these in Appendix \ref{sec:artefacts}.

\begin{figure}
    \includegraphics[width=\columnwidth]{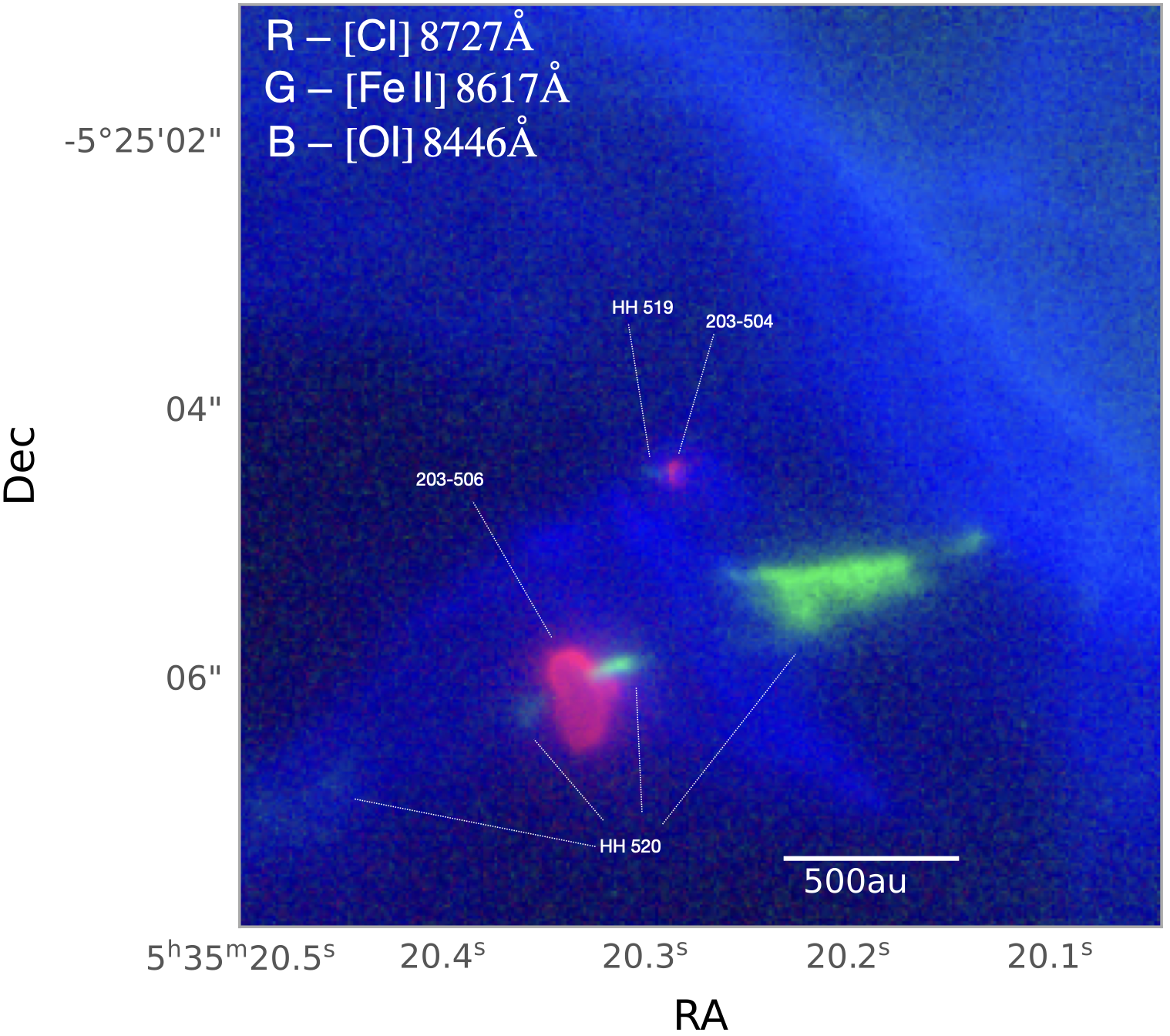}    
    \caption{A 3 colour composite image from our dataset consisting of [C\,\textsc{i}] 8727 \AA\ (red), [Fe~\textsc{ii}] 8617 \AA\ (green) and [O\,\textsc{i}] 8446 \AA\ (blue). The [Fe~\textsc{ii}] traces the bipolar outflow, HH 520, from 203-506 and the compact outflow HH 519 from 203-504.   }
    \label{fig:MUSEoverview2}
\end{figure}

\begin{figure*}
    \vspace{-0.4cm}
    \hspace{-0.4cm}
    \includegraphics[width=1.99\columnwidth]{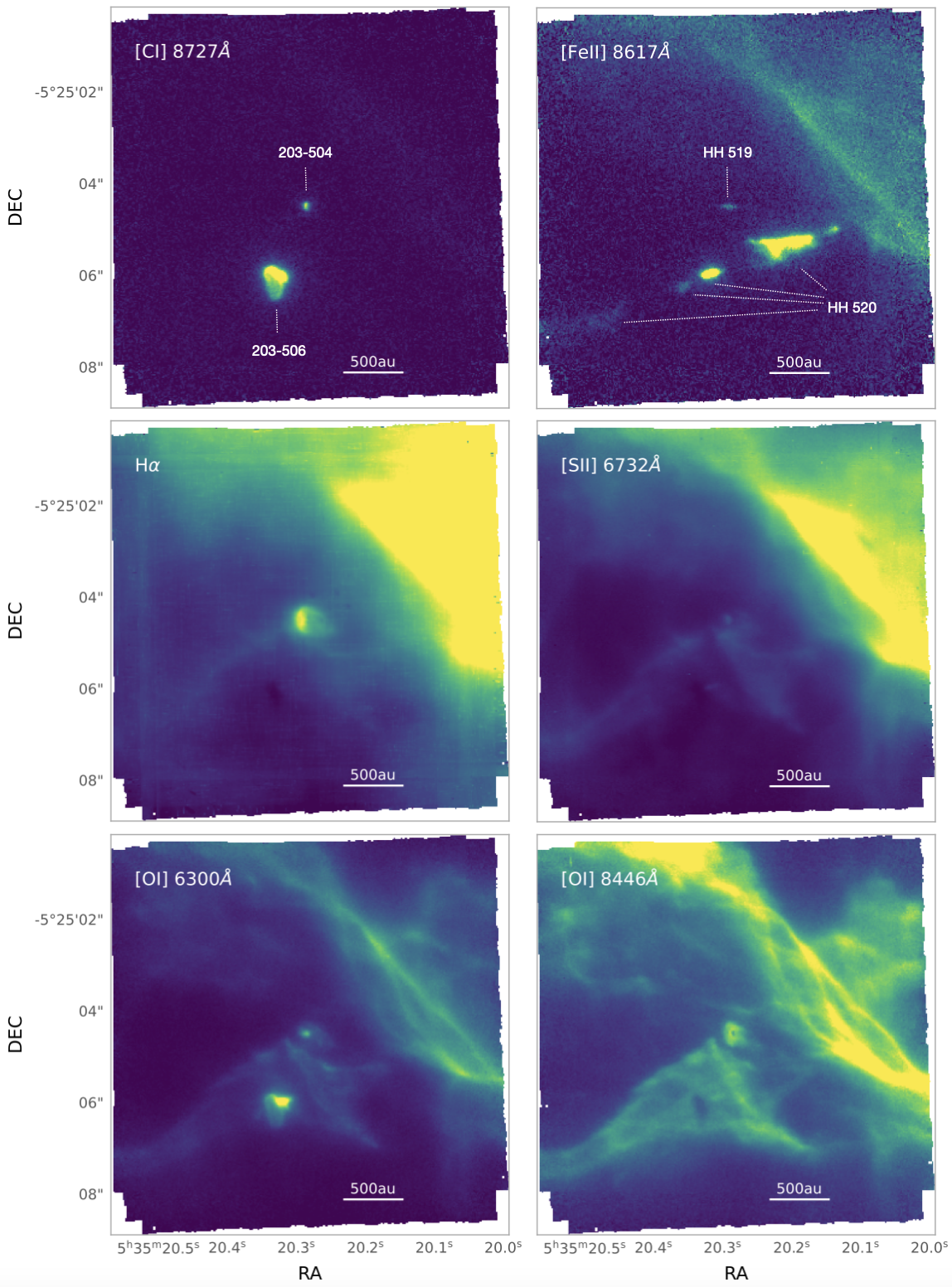}
    \caption{A gallery of some key continuum-subtracted lines towards 203-504/203-506 in the Bar. The upper left panel is [C\,\textsc{i}], which traces FUV irradiated material in the discs. The upper right panel is [Fe \,\textsc{ii}], which traces shock excited gas from jets/outflows. Centre-left and centre-right are H$\,\alpha$ and [S\,\textsc{ii}]  which trace EUV-irradiated gas. The lower panels are [O\,\textsc{i}] 6300 \AA\ (left) which can trace collisionally and shock excited gas as well as regions of OH dissociation, and the [O\,\textsc{i}] 8446 \AA\ line (right) which traces Lyman $\beta$ irradiated gas.}
    \label{fig:gallery}
\end{figure*}

\begin{table*}
    \centering
    \begin{tabular}{cccc}
    \hline
    Line & Wavelength & Primarily Traces & Notes   \\
        &  (\AA)   & &   \\
    \hline    
     H\,$\beta$& 4861 & The H\,\textsc{\,\textsc{ii}} region. Proplyd 203-504. The irradiated Delta Feature. \\ 
     $\left[\textrm{O\,\textsc{\,\textsc{ii}i}}\right]$& 5007 & The H\,\textsc{\,\textsc{ii}} region.  \\     
     $\left[\textrm{N\,\textsc{ii}}\right]$& 5755 & The H\,\textsc{\,\textsc{ii}} region. Proplyd 203-504. The irradiated Delta Feature. \\      
     $\left[\textrm{OI}\right]$& 6300 & HH 519. 203-506 disc and inner part of HH 520.  Delta Feature and Bar PDR.      \\
     $\left[\textrm{S\,\textsc{\,\textsc{ii}i}}\right]$& 6312 & The H\,\textsc{\,\textsc{ii}} region. Proplyd 203-504 (faint). \\           
     $\left[\textrm{O\,\textsc{i}}\right]$& 6363 & HH 519. 203-506 disc and inner part of HH 520.  Delta Feature and Orion Bar PDR.      \\     
     H\,$\alpha$& 6582.8 & The H\,\textsc{\,\textsc{ii}} region. Proplyd 203-504. The irradiated Delta Feature. & 203-506 in silhouette \\                
     He\,\textsc{i} &6678 & The H\,\textsc{\,\textsc{ii}} region. Proplyd 203-504 & 203-506 in silhouette  \\     
     $\left[\textrm{N\,\textsc{ii}}\right]$& 6548 & The H\,\textsc{\,\textsc{ii}} region. Proplyd 203-504. The irradiated Delta Feature. & \\         
     $\left[\textrm{N\,\textsc{ii}}\right]$& 6583 & The H\,\textsc{\,\textsc{ii}} region. Proplyd 203-504. The irradiated Delta Feature. \\    
     $\left[\textrm{S\,\textsc{ii}}\right]$& 6716 & The H\,\textsc{\,\textsc{ii}} region. Part of HH 520 & 203-506 in silhouette \\        
     $\left[\textrm{S\,\textsc{ii}}\right]$& 6732 & The H\,\textsc{\,\textsc{ii}} region \\    
     He\,\textsc{i} &7065 & The H\,\textsc{\,\textsc{ii}} region. Proplyd 203-504 & 203-506 in silhouette  \\          
     $\left[\textrm{Fe \,\textsc{ii}}\right]$& 7155 & The outflows, HH 519 and HH 520 \\         
     $\left[\textrm{Fe \,\textsc{ii}}\right]$& 7172 & The outflows, HH 519 and HH 520 \\              
     $\left[\textrm{O\,\textsc{i}}\right]$& 7255 & 203-504 proplyd.  Delta Feature and Bar PDR. & 203-506 in silhouette\\     
     He\,\textsc{i} &7282 & The H\,\textsc{\,\textsc{ii}} region. Proplyd 203-504 & 203-506 in silhouette  \\          
     $\left[\textrm{Ca \,\textsc{ii}}\right]$& 7291 & The outflows, HH 519 and HH 520 \\        
    $\left[\textrm{Ar \,\textsc{\,\textsc{ii}i}}\right]$    & 7751    & The H\,\textsc{\,\textsc{ii}} region. Proplyd 203-504     & 203-506 in silhouette \\       
    $\left[\textrm{O\,\textsc{i}}\right]$& 8446 & 203-504 proplyd.  Delta Feature and Bar PDR. & 203-506 in silhouette\\ 
     $\left[\textrm{Fe \,\textsc{ii}}\right]$& 8617 & The outflows, HH 519 and HH 520 \\    
    $\left[\textrm{C}\,\textsc{i}\right]$ & 8727 & Disc/wind of 203-504 and 203-506  \\
     $\left[\textrm{S\,\textsc{\,\textsc{ii}i}}\right]$& 9069 & The H\,\textsc{\,\textsc{ii}} region. Proplyd 203-504. \\        
    \hline     
    \end{tabular}
    \caption{A non-exhaustive summary of lines detected in our MUSE dataset, and the features they trace. }
    \label{tab:lines}
\end{table*}

\subsection{Two extremely different photoevaporating discs}
\label{sec:proplyds}
Our FOV contains two protoplanetary discs subject to external photoevaporation. Their projected on-sky separation is only around 660\,au, but the nature of their external photoevaporation is considerably different and they are probably not located physically close to one another.

\subsubsection{203-504}
\label{sec:504}
203-504 has a classic teardrop shaped proplyd morphology that is clear in ionisation front tracers such as H\,$\alpha$. The 203-504 proplyd points towards $\theta^2$ Ori A, implying that is the UV source predominantly responsible for driving the mass loss \citep[as already pointed out by][]{2017ApJ...837..151O}. In addition, there is [C~{\sc i}] 8727 \AA\ and [O\,\textsc{i}] 8446 \AA\ emission from FUV/Lyman $\beta$ irradiated gas (see Figure \ref{fig:gallery}). 203-504 is hence being exposed to both FUV and EUV radiation.

In ionisation equilibrium, the extent of the ionisation front in a proplyd depends on the ionising flux incident upon the proplyd and the integrated recombination rate from the UV source down to the I-front. Assuming that the flow diverges spherically from the I-front and that a spherical flow from the disc into the I-front has to satisfy mass conservation, one can estimate the required mass loss rate to place the I-front at its observed radius. For example, for a given incident ionising flux, decreasing the mass-loss rate would make the ionisation front radius smaller \citep[e.g.][]{1998ApJ...499..758J, 2015MNRAS.446.2944C}. Specifically, for a proplyd at a projected separation $d$ from a UV source producing ionising photons at a rate $\dot{N}_{ly}$ per second and with mass-loss rate $\dot{M}_{\textrm{wind}}$, the ionisation front radius is
\begin{multline}
    R_\mathrm{IF} \approx 1200 \, \left( \frac{\dot{M}_\mathrm{wind}}{10^{-8} \, M_\odot \, \mathrm{yr}^{-1}}\right)^{2/3} \\ 
    \times \left( \frac{\dot{N}_\mathrm{Ly}}{10^{45} \, \mathrm{s}^{-1}}\right)^{-1/3} \left( \frac{d}{1\,\mathrm{pc}} \right)^{2/3} \, \mathrm{au}
    \label{equn:RIFMdot}
\end{multline} 
\citep{2022EPJP..137.1132W}.

The projected separation from 203-504 to $\theta^2$ Ori A is $d=0.077$\,pc assuming a distance of 400\,pc. For the ionising luminosity of $\theta^2$ Ori A we use the value from Table 2.3 of \cite{2006agna.book.....O} of $3.63\times10^{48}$\,photons\,s$^{-1}$. We measure an ionisation front radius for 203-504 of 139\,au, which combined with the values above corresponds to a mass loss rate of $\sim3.1\times10^{-7}$\,M$_\odot$\,yr$^{-1}$. In making this estimate we have assumed based on the I-front morphology that only $\theta^2$ Ori A is driving the mass loss. We will discuss the possible relative locations of the evaporating discs/UV sources and the ISM in section \ref{sec:3dstructure}. 

As discussed, the EUV field sets the ionisation front radius, but the FUV radiation is expected to be what actually drives the wind launching and sets the mass-loss rate \citep[][Haworth et al. submitted]{1998ApJ...499..758J, 2004ApJ...611..360A, 2018MNRAS.481..452H}. Assuming that the FUV incident upon 203-504 is also dominated by  $\theta^2$ Ori A and that  $\theta^2$ Ori A and 203-504 are the same distance from the observer, integrating over a \cite{1979ApJS...40....1K} model spectrum suggests an FUV field incident upon 203-504 of approximately $8\times10^4$G$_0$. Other proplyds at a comparable FUV field strength are seen for in NGC 2024, which are irradiated by an O8V star and also have mass loss rates of order $10^{-7}$\,M$_\odot$\,yr$^{-1}$ \citep{2021MNRAS.501.3502H}. The majority of proplyds in the ONC are in the vicinity of $\theta^1$ Ori C irradiated by $\geq 10^6$\,G$_0$ UV fields with correspondingly higher mass loss rates (regularly $\geq10^{-6}$\,M$_\odot$\,yr$^{-1}$). 

\subsubsection{203-506}
\label{sec:506}

The other young stellar object, 203-506 has a drastically different morphology, not seen in any other proplyds. The main feature that is unusual in this system is that it has no ionisation front, evidence that it is not being irradiated by EUV photons. It is however being irradiated by the FUV, which is responsible for driving the [C\,\textsc{i}] 8727\,\AA\ (see Appendix \ref{sec:appCi}) and the \oi\  6300\,\AA\ line by OH dissociation (see section \ref{sec:OI}).

203-506 is bright in \oi\ 6300 \AA\ and \Ci\ 8727 \AA , but appears in silhouette in \Ha , \oiii\ 5007 \AA , \nii\ 6583 \AA , \oi\ 8446 \AA , and \sii\ 6732 \AA . Reference to images made immediately after installation of the WFPC2 on the HST show the same properties in the superior \Ha\ F656N and \nii\ F658N images. The fact that the silhouettes are similar in all 
ions (except when 203-506 is in emission) means that 203-506 lies well into the foreground of the MIF.

\subsection{Jets/outflows}
\label{sec:jetsoutflows}
Our FOV contains two Herbig-Haro (HH) objects: HH 519 and HH 520 associated with 203-504 and 203-506 respectively. They were originally identified by \cite{2000AJ....119.2919B} as a pair of oppositely directed monopolar outflows in narrowband [O\,\textsc{i}] 6300\,\AA\ images, as well as H\,$\alpha$ in the case of HH 519. 

 In our data,  HH 520 (emanating from 203-506) is clearly observable in jet tracing \citep{2013MNRAS.433.2226R, 2015MNRAS.450..564R, 2016MNRAS.463.4344R,2023A&A...673A.166K} [Fe II] emission, specifically the 8617 \AA\ a$^4$F$_{9/2}$-a$^4$P$_{5/2}$ line. Of note is that we also detect fainter, oppositely directed, [Fe \,\textsc{ii}] emission from HH 520 (see Figure \ref{fig:MUSEoverview2} and the upper right panel of Figure \ref{fig:gallery}). MUSE reveals the eastern limb of the jet for the first time, demonstrating that this is a bipolar flow, but with the south eastern component a factor 5-10 fainter than the previously identified north western component. 

\begin{figure}
    \centering
    \includegraphics[width=\columnwidth]{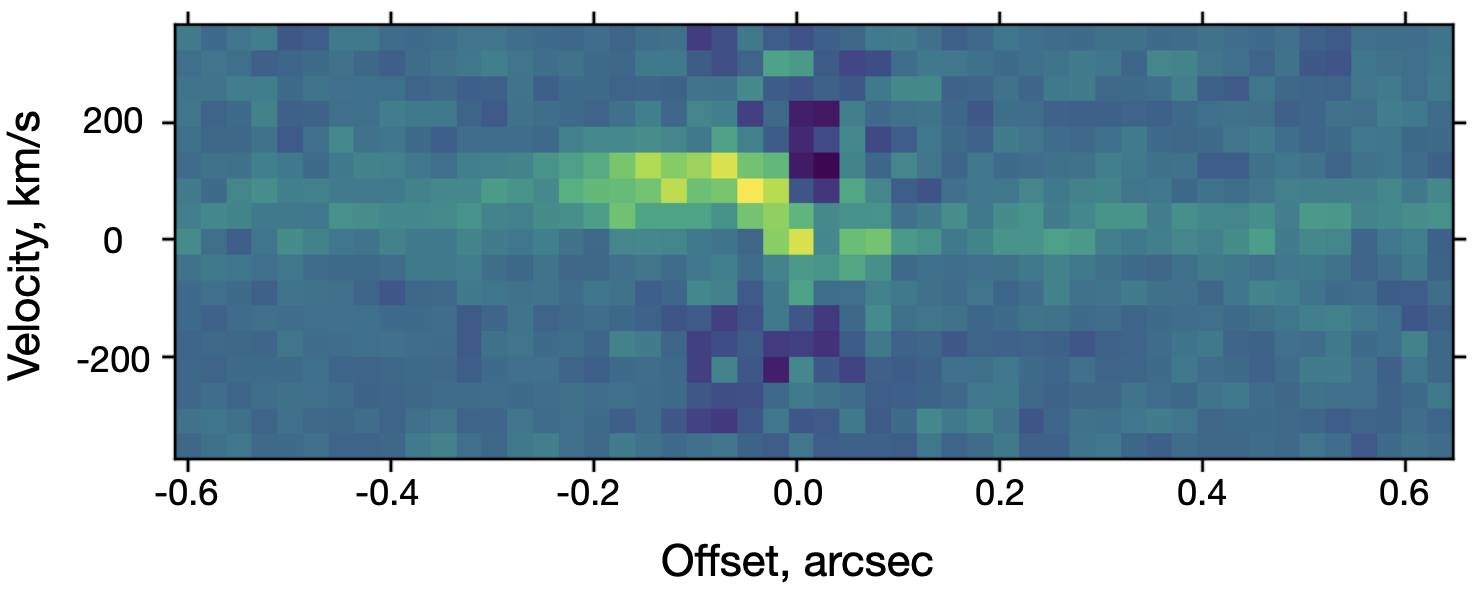}

    \vspace{-0.2cm}
    \hspace{-1.5cm}
    \includegraphics[width=1.2\columnwidth]{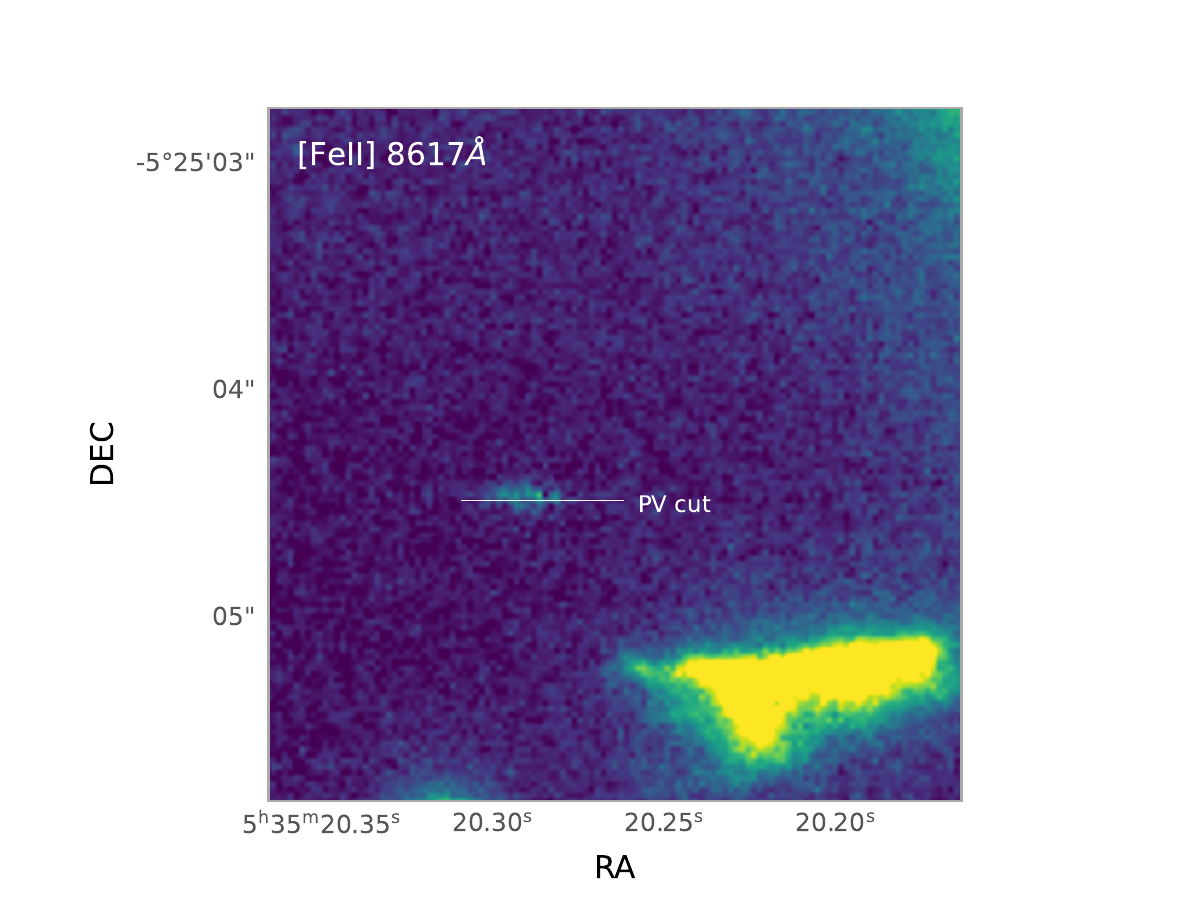} 
    \caption{A continuum-subtracted position-velocity diagram in a cut across the YSO 203-504, with the reference velocity corresponding to the [Fe \,\textsc{ii}] 8617 \AA\ line. The velocity of the outflow HH 519 is resolved at $\sim130\,$km\,s$^{-1}$. The lower panel shows the location of the cut (compare with the upper right panel of Figure 5).}
    \label{fig:FeIIPV203504}
\end{figure}

Our data for HH 519 (for example in Figure \ref{fig:gallery}) demonstrates that the H\,$\alpha$ previously identified as the outflow actually traces the northeast bright rim of the Delta Feature. The underlying cloud is particularly clear in the [O\,\textsc{i}] 6300 \AA\ and [O\,\textsc{i}] 8446 \AA\ emission in the lower panels of Figure \ref{fig:gallery}. 203-504 is coincidentally located close to the northern vertex of that irradiated Delta Feature along the line of sight, making it appear as though the elongated bright rim originated from it. However, the high angular resolution of MUSE NFM (0.0921'') also demonstrates that 203-504 is completely disconnected from the irradiated Delta Feature (particularly in [O\,\textsc{i}] 8446\AA\,). 

Our data also reveal a compact and kinematically resolved high velocity outflow feature in [Fe \,\textsc{ii}] 8617\AA\ associated with 203-504. Because the outflow is associated with the same YSO the name ``HH 519'' is retained for this new outflow associated with 203-504 (it is labeled in Figure \ref{fig:MUSEoverview2} and in the upper right panel of Figure \ref{fig:gallery}). In Figure \ref{fig:FeIIPV203504} we make a cut (a single pixel wide) across the HH 519 jet axis in continuum subtracted [Fe \,\textsc{ii}] emission and show the position-velocity diagram. The outflow appears to be bipolar, with each outflow limb extending $\sim0.4''$ and line of sight velocity $\sim130\,\textrm{km}\,\textrm{s}^{-1}$ that is consistent with velocities measured in other HH jets \citep[e.g.][]{2001ApJ...559L.157H, 2005AJ....130.2197H, 2007ApJ...660..426H, 2017MNRAS.470.4671R, 2022MNRAS.517.5382R,2023A&A...673A.166K}. The compact size and fact that it is spectrally resolved suggests that HH 519 is tilted toward the line of sight by $\gtrsim45^{\circ}$.

\section{Discussion}

\subsection{The Irradiation of our FOV}

\subsubsection{Strong Line ratios argue for irradiation by \tC\ and/or \tX}
\label{sec:Strong}
\citet{2017ApJ...837..151O} used line ratios to determine the source of UV irradiation at the MIF across the Huygens region, and were able to distinguish when $\theta^1$ Ori C and $\theta^2$ Ori A were the dominant source (with one field that they studied likely receiving significant contributions from both sources). To compare with  \citet{2017ApJ...837..151O}, three small sub-samples were taken. Within the Delta Feature the sub-sample was 0\farcs52$\times$0\farcs52 centred 0\farcs78 east and 0\farcs15 south of 203-506. Outside of the Delta Feature an east sub-sample was 0\farcs69$\times$1\farcs25 centred 2\farcs38 east and 0\farcs37 north of 203-506 and a west sub-sample of  0\farcs80$\times$0\farcs61 centred  1\farcs39 west and 0\farcs07 north of 203-506. The averages of these three sub-samples are shown in Table~\ref{tab:ratios}, and compared also with the other measurements in the Huygens Region. The values in our field lie between those in the central Huygens Region and those to the SE of the Bar, as one would expect given our field lies between the two. \citet{2017ApJ...837..151O} determined that their sample E (OKF-E) is likely subject to significant irradiation by both $\theta^2$ Ori A and $\theta^1$ Ori C. Our FOV is most similar to OKF-E, suggesting both stars may also play a role here, though with perhaps a fractionally more significant contribution from $\theta^1$ Ori C. 

\begin{table}
\caption{Huygens Region and Delta Line Ratios. OKF refers to a sample presented in \citet{2017ApJ...837..151O}}
\label{tab:ratios}
\begin{tabular}{lcc}
\hline
\hline
{Sample}&{I([O\,\textsc{iii}] 5007)/I(\Hb)}&{I([N\,\textsc{ii}]\,6583)/I(\Ha)} \\
\hline
\ \cite{2004MNRAS.355..229E}$^{a}$   &     3.84            &   0.13          \\
\ OKF-A$^{a}$          &     3.73            &   0.12           \\
\ OKF-B $^{a}$         &     3.58            &   0.14           \\
\ Ave 203-506$^{b}$  &     2.16$\pm$0.04   &   0.23$\pm$0.01   \\
%\ OKF-D $^{c}$         &     1.06            &   0.31           \\
\ OKF-E$^{c}$          &     1.96            &   0.18            \\
\hline
\end{tabular}\\
~$^{a}$Samples near \tC.\\
$^{b}$Average of Delta Feature and two adjacent samples.\\
$^{c}$Sample in \citet{2017ApJ...837..151O} nearest 203-504/6, which they determined was affected by both $\theta^2$ Ori A and $\theta^1$ Ori C.
\end{table}

\subsubsection{Comparison of line ratios over a wide range of ionization energies argue for both \tC\ and \tX }
\label{sec:range}

We additionally compared the average line fluxes over a bigger range of ionization in our samples described above to those in the vicinity of $\theta^1$\,Ori C in Table \ref{tab:fluxes}. The Inner Delta values are from the ``Ave 203-506'' samples in Table~\ref{tab:ratios} and the Average Background samples are from outside the Delta Feature. These fluxes are defined relative to $F($H$\,\beta)=100$.  We see that the high ionisation [O\,\textsc{iii}] 5007\,\AA\ line is much weaker than in the vicinity of $\theta^1$ Ori C, while the intermediate ionisation [N\,\textsc{ii}] 6853\,\AA\ line is stronger and the low ionisation [O\,\textsc{i}] 6300\,\AA, [S\,\textsc{ii}] 6732\,\AA\ and [O\,\textsc{i}] 8446\,\AA\ lines are all much stronger. This is consistent with the known pattern of decreasing ionization with increasing distance from \tC\ \citet{2010AJ....140..985O}. The average background and fluxes in the inner part of the Delta Feature are similar, with the exception of the [O\,\textsc{i}] 8446\,\AA\ line, which is elevated by over a factor 2 compared to the background and by over a factor 3 compared to the vicinity of $\theta^1$ Ori C, which as we discuss in more detail in appendix \ref{sec:appveldif}.

Both this section and its predecessor both point towards the radiation field in our data not being exceptional compared to other samples in the region, i.e. that both stars can play a role in irradiating our FOV.

The presence of the distinct boundaries on the Delta Feature therefore point to genuine structure, such as density enhancements and/or a raised clump in the shoulder region of the Bar's MIF.

\begin{table*}
\caption{Line fluxes near $\theta^1$ Ori C, ''outside'' and ``inside'' the Delta Feature. The fluxes are defined relative to the H$\,\beta$ line with a value of 100. }
\label{tab:fluxes}
\begin{tabular}{lccc}
\hline
\hline
{Ion,Line}                                    &{Near \tC$^{a}$}&{Average Background}&{Delta Feature} \\
\hline
\ \Hb                                            &           100          &    100                              &    100              \\
\ [O\,\textsc{iii}] ,5007 \AA           &           384           &   216                               &    220              \\
\ [O\,\textsc{i}] ,6300 \AA             &           0.71          &    1.0                               &      1.6              \\
\ \Ha                                            &            287          &      293                             &    293         \\
\ [N\,\textsc{ii}] , 6583 \AA           &            38             &     70                               &      65             \\
\ [S\,\textsc{ii}] , 6732 \AA           &            3.5             &   7.8                               &      8.6                 \\
\ [O\,\textsc{i}] , 8446 \AA           &             0.88           &   1.3                              &       2.8                \\
\hline
~$^{a}$From \citet{2004MNRAS.355..229E}.
\end{tabular}\\
\end{table*}

\subsubsection{Information from the Delta Feature}
\label{sec:DeltaFeatures}
 The Delta Feature has three edges, one in the north east, one in the north west and one in the south. The NE and NW boundaries of the Delta Feature are more sharply
defined than the south boundary. They have the appearance of ionization fronts
viewed edge-on and are bright in [O\,\textsc{i}] 6300\,\AA, [S \textsc{ii}] 6732\,\AA\, [N\,\textsc{ii}] 6583\,\AA, and [O\,\textsc{i}] 8446\,\AA. In H$\alpha$ and H$\beta$ they are more diffuse, but are bright and in about the same positions. None of the edges of the Delta Feature are associated with changes in [C\,\textsc{i}] 8727\,\AA\ or [O\,\textsc{iii}] 5007\AA. These characteristics all point towards the rim of the Delta Feature being irradiated by some EUV radiation with no significant radiation above 24.6\,eV (due to the lack of the helium ionised zone tracing [O\,\textsc{iii}]). The NW boundary is approximately perpendicular to a line towards $\theta^2$\,Ori A and the NW boundary is approximately perpendicular to a line towards $\theta^1$ Ori C, suggesting that the NE and NW edges of the Delta Feature are being irradiated by the two different UV sources. This is consistent from the two preceding sections concluding that our FOV is likely to be influenced by both \tC\ and \tX . The presence of the distinct boundaries on the Delta Feature point to genuine structure, such as a density enhancement and/or a raised clump in the shoulder region of the Bar's MIF.

\subsection{The locations of the externally photoevaporating discs }
\label{sec:3dstructure}

\begin{figure}
    \centering
    \includegraphics[width=\columnwidth]{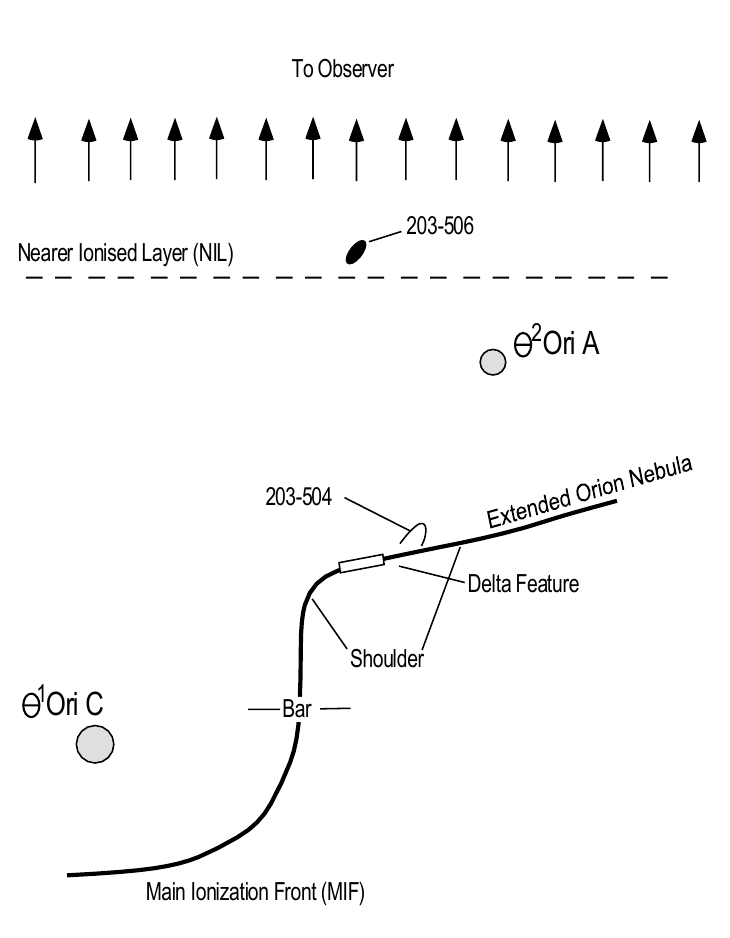}
    \caption{A cartoon schematic (not to scale) of the Huygens region and the key features in our field. $\theta^1$ Ori C sits within a concave surface. Where that surface orients  along the line of sight is the Bar ionisation front. Our field, with the Delta Feature and the two evaporating discs, 203-504 and 203-506, lies in the direction of the shoulder region, where \tC 's radiation begins to be shadowed. The nearer ionised layer (NIL) is somewhere in the foreground of $\theta^1$ C but the exact distance is not well known, particularly near $\theta^2$ Ori A \protect\citep{2020ApJ...891...46O}.} 
    \label{fig:profile}
\end{figure}

In section \ref{sec:proplyds} we discussed that 203-504 and 203-506, with a projected separation of 660\,au on the sky, appear to be being subjected to external photoevaporation by very different external UV radiation fields. 203-504 is being irradiated by both direct EUV and FUV radiation, whereas 203-506 is solely being irradiated by the FUV and is hence somehow shielded from the EUV from $\theta^2$ Ori A and $\theta^1$ Ori C. Here we discuss the possible structure of the region that could give rise to this scenario. To do this, we can draw on existing information about the structure of the nebula from the Huygens Region, past the Bar, and into the Extended Orion Nebula (EON). 

\subsubsection{Location of key stars and features that block portions of their radiation}

Figure~\ref{fig:profile} shows a cartoon schematic cross section from near \tC\ through the region of the Bar that contains our FOV and outward into the EON. Not all objects lie in the same direction from \tC, so this two dimensional sketch is accurate only in the sense of the pattern of irradiation discussed in this section. 
The smooth line represents the MIF. \citet{2017ApJ...837..151O} most recently established that inside (closer to \tC ) the Bar the MIF 
is created by EUV from \tC .
As the MIF tilts toward the observer limb brightening increases the observed surface brightness until reaching a maximum along the MIF's ionization front. Beyond that the MIF begins to flatten, although receiving enough EUV from \tC\ and later \tX\ to make the MIF bright. This broad region immediately across the line of maximum tilt is usually called the Bar (not simply the narrow maximum tilt feature).  This model of the structure of the Bar has been the result of low velocity resolution spectra that explain the observed ionization changes \citet{2011MNRAS.417..420M}, and several high velocity resolution studies employing long-slit spectra parallel to the Bar made in sequences crossing the Bar \citep{2017ApJ...837..151O,2018MNRAS.478.1017O}. Of course beyond the MIF lies a PDR, that in the section just outside the visual peak of brightness has produced a cottage-industry output of insightful infrared and radio studies. 

At the outer parts of the shoulder EUV from \tX\ rapidly begins to be dominant, a condition that increases when moving to greater distances from \tC\ and \tX . The rapid transition at the shoulder indicates that the shoulder is blocking \tC 's EUV from reaching the MIF further out, with the consequence that \tX\ becomes the most important star when looking beyond the vicinity of the Bar's shoulder \citep{2011MNRAS.410.1320R,2017ApJ...837..151O}.  

In their spectroscopic study of the Huygens Region \citet{2007AJ....133..952G} found an extended low ionization layer of gas northwest of the Bar, blueshifted with respect to the MIF. Subsequent studies in other emission-lines \citet{2019ApJ...881..130A} established that this was an ionized layer that in the vicinity of the Trapezium was about 0.4 pc closer to the observer than \tC\ \citep{2020ApJ...891...46O}. Although called several names in early studies, the most recent \citep{2020ApJ...891...46O} name is the nearer ionized layer (NIL), a name indicating that it is closer to the observer than the MIF that is located on the far side of \tC . The source of the ionization  in the center of the Huygens Region is certainly \tC\ while further out \tX\ probably plays an important role. One sees the NIL in spectra of the shoulder region. In the central Huygens Region, the NIL velocity is about 6$\pm$2 \kms\ \citep{2020ApJ...891...46O} 
changing to about 0$\pm$2 \kms\ beyond the Bar \citep{2017ApJ...837..151O,2018MNRAS.478.1017O}.  The change of velocity in the outer region probably indicates that the distance of the NIL is influenced by the rising slope of the MIF.

The average velocity of the NIL emission-lines is similar to the \heI\ 3889 \AA\ absorption line seen in the Trapezium stars (average 2.1$\pm$0.6 \kms) and in \tX\  (two components, at -2.9 \kms\ and 5.2 \kms ). The 2$\rm ^{3}S$ lower level of the 3889 \AA\ line is a metastable level populated by recombinations of ionized He and this absorption line most likely arises from the ionized NIL. Since the 3889 \AA\ absorption line is present in the spectrum of \tX, the NIL must lie on the observer's side of the star. Since the NIL is
ionization bounded (optically thick to LyC radiation) \citep{2019ApJ...881..130A} this means that the NIL permits only \tX 's FUV to pass. 

\subsubsection{Constraints on the location of 203-504}

The bright rim of the teardrop shaped 203-504 is clearly pointed towards \tX , as first shown in \citet{2000AJ....119.2919B}. Seen in emission in multiple ions, it shows no evidence of influence by \tC\ and the EUV of that star must be blocked by intervening gas. These conditions are satisfied by the proplyd being close to the MIF at the more distant side of the shoulder, where 
\tC 's radiation is blocked by the curved PDR that lies underneath. This is shown in Figure~\ref{fig:profile}. 

\subsubsection{Constraints on the location of 203-506}

As discussed in Section~\ref{sec:506} 203-506 is irradiated only by FUV, although it is not clear from which nearby star. As shown in Figure~\ref{fig:profile}, this requires it to be located on the observer's side of the NIL. This foreground location is shared by the large number of silhouette-only proplyds seen within the Huygens Region. These were originally thought to be a product of blocking by the foreground Veil neutral gas \citep{1994ApJ...436..194O}, but with the recognition of the NIL, it now seems the latter is the more likely blocking agent. 

The MIF is much higher surface brightness than the NIL, which means that the fact that 203-506 is seen in silhouette indicates that it is well into the foreground from the MIF but says nothing about its position relative to the NIL.

\subsection{Atomic carbon associated with 203-506. }
\label{sec:extPhotoCI}
External photoevaporation is typically identified based on the proplyd teardrop shaped ionisation front morphology, especially in lines like H\,$\alpha$ and Pa\,$\alpha$ \citep{2000AJ....119.2919B, 2008AJ....136.2136R, 2016ApJ...826L..15K, 2021MNRAS.501.3502H}. To understand the full extent to which external photoevaporation operates we need to be able to identify it even when there isn't a clear teardrop shaped ionisation front. Specifically, when the wind is weaker, or when a disc is irradiated only by an FUV radiation field. \cite{2020MNRAS.492.5030H} used PDR-dynamical models and synthetic observations to study the observational characteristics of winds without ionisation fronts. They found that CO is typically not expected to be a good tracer of external photoevaporative winds because it is dissociated near the base of the flow, making it hard to distinguish from a Keplerian disc. Conversely, \cite{2020MNRAS.492.5030H} predicted that C\,\textsc{i} emission could make a good kinematic tracer of the inner FUV driven parts of winds, however that has not been demonstrated in practice yet. \cite{2022MNRAS.512.2594H} targeted the C\,\textsc{i} $^3$P$_1$-$^3$P$_0$ 492.16\,GHz line towards proplyds in NGC 1977 with APEX, but obtained no clear detections. The most likely explanation for this is thought to be low disc masses, but it could also be due to carbon depletion or beam dilution. 

203-506 is the first known instance of an FUV-only externally driven wind from a protoplanetary disc with no ionisation front \citep[even the weakly irradiated 114-426 has an extended H\,$\alpha$ ionisation front][]{2012ApJ...757...78M}. In the context of the above discussion, the detection of the [C\textsc{i}] 8727\AA\ line towards 203-506 illustrated in Figure \ref{fig:MUSEoverview} is the first clear demonstration that carbon is abundant in the FUV driven part of the wind (see Appendix \ref{sec:appCi} for discussion on the driving mechanism of that line). Although the spectral resolution of MUSE is inadequate to measure the expected wind velocity ($\sim4$\,km\,s$^{-1}$), this makes 203-506 the most promising known target for validating the kinematic ALMA predictions for C\,\textsc{i} winds by \cite{2020MNRAS.492.5030H}. Also of note is that more compact [C\,\textsc{i}] 8727\AA\ emission is detected off-centre of the YSO and interior to the I-front of 203-504 (see the upper left panel of Figure \ref{fig:gallery}).  We estimate [C\,\textsc{i}] 8727\,\AA\ luminosities of $1.1\times10^{-5}$~L$_\odot$ and $6.4\times10^{-7}$~L$_\odot$ for 203-506 and 203-504 respectively. 

Goicoechea et al. (priv. communication) have a spatially unresolved [C\,\textsc{i}] 3P$_1$-3$P_0$ detection with ALMA towards 203-506, finding a peak flux density of 500\,mJy/beam at $0.5''$ resolution. This makes 203-506 the best known target for high spatial and spectral resolution follow up with ALMA to study the kinematics of the FUV heated component of external photoevaporative winds.

\begin{figure*}
    \centering
   \includegraphics[width=1.8\columnwidth]{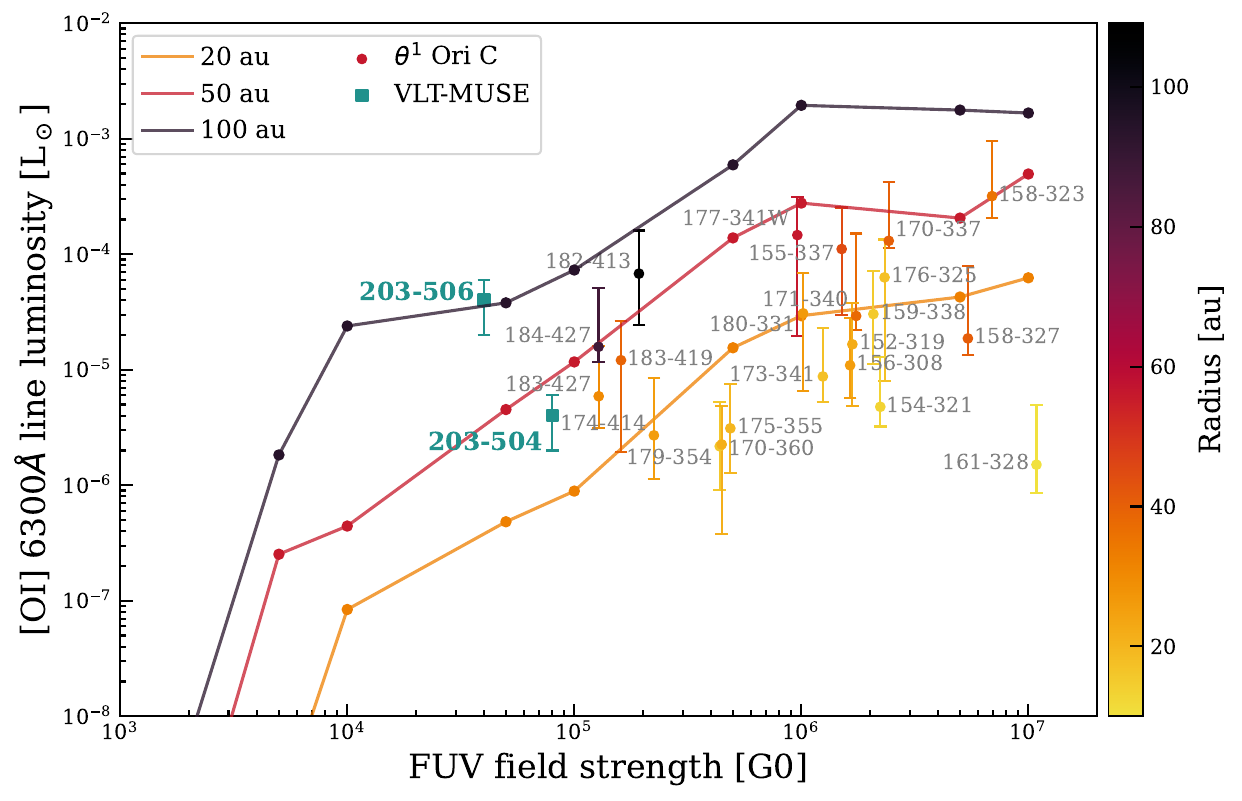}
    \caption{The lines are \protect\cite{2023MNRAS.518.5563B} model predictions of the [O\,\textsc{i}] 6300\AA\ luminosity due to external photoevaporation as a function of external FUV radiation field strength and disc radius. The majority of points are data for ONC proplyds \protect\citep{2023MNRAS.518.5563B}. The square green points are our new measurements for 203-504 and 203-506. }
    \label{fig:BallabioPlot}
\end{figure*}

\subsection{[O\,\textsc{i}] 6300\AA\ emission from 203-504 and 203-506}
\label{sec:OI}

Outside of the extreme conditions in our two targeted proplyds, \oi\ 6300\,\AA\ radiation follows collisional excitations from the ground-state $\rm ^{3}P$ levels to the $\rm ^{1}D_{2}$ state 2.0 eV above it. This accounts for the appearance of this line in the MIF and the edge-on view in the Bar. However \cite{1998ApJ...502L..71S} showed that [O\,\textsc{i}] 6300\AA\ can also be emitted following the FUV driven photodissociation of OH. This was recently generalised by \cite{2023MNRAS.518.5563B} who demonstrated that theoretical models of [O\,\textsc{i}] 6300\AA\ emission by that mechanism are consistent with the [O\,\textsc{i}] luminosities from the PDR (i.e. excluding the contribution from the I-front and jets) of a large number of ONC propylds and predict that the line could be used to identify external photoevaporation in distant clusters. A continuum subtracted map of the [O\,\textsc{i}] 6300\AA\ emission from our FOV is given in the lower left panel of Figure \ref{fig:gallery}. 

One of the issues with studying the [O\,\textsc{i}] 6300\AA\ line is that it can originate from multiple sources, including the PDR (as described above), thermally from internal winds/an ionisation front and as part of high velocity outflows. {With our data we are unable to spectrally distinguish any low and high velocity components, but do spatially resolve the discs and outflows so can try to avoid them that way}. In the case of 203-504 we estimate the [O\,\textsc{i}] line luminosity using an annulus that excludes the bright central component associated with the outflow and the star itself. For 203-506 we exclude the bright spot on the northwest side of the disc in the vicinity of the outflow. Taking this approach we obtain [O\,\textsc{i}] 6300\AA\ luminosities of $4\pm2\times10^{-6}$L$_\odot$ and $4\pm2\times10^{-5}$\,L$_\odot$ for 203-504 and 203-506 respectively. We plot these alongside the models and other ONC proplyd data from \cite{2023MNRAS.518.5563B} in Figure \ref{fig:BallabioPlot}. Both discs have [O\,\textsc{i}] luminosities consistent with the models. 203-504 has a luminosity similar to the other ONC proplyds. 203-506 is brighter than many of the proplyds, {and at the high end of the observed range of luminosities found by \citep{2018A&A...609A..87N},} but agrees well with the model prediction for a 100\,au disc (which is the observed radius of 203-506). That 203-506 is so bright in [O\,\textsc{i}] 6300\AA\ further supports the interpretation of \cite{1998ApJ...502L..71S} and \cite{2023MNRAS.518.5563B} that [O\,\textsc{i}] 6300\AA\ is produced in the PDR following the dissociation of OH and not just thermally in ionised gas or in the shocks of outflows. Zanesse et al. (in prep.) also use marginally resolved JWST observations to demonstrate that 203-506 has a large OH abundance. 

\vspace{-0.2cm}
\section{Summary and conclusion}

We use VLT-MUSE NFM to study two externally photoevaporating discs and their outflows, situated towards Orion's SW-Bright Bar.  We draw the following main conclusions from this work. \\

\noindent 1) Based on emission line fluxes and ratios, our field is consistent with the wider irradiation structure of the surrounding ONC. It shows no evidence of higher ionisation as seen in the vicinity of $\theta^1$ Ori C and sits in a regime where it is possible that both $\theta^1$ Ori C and $\theta^2$ Ori A are contributing at a significant level to the radiation field in the region. Our field contains a $\Delta$ shaped ``Delta Feature'' which is a structure in the Bar that exhibits relatively sharp boundaries towards the NE (in the direction of $\theta^2$ Ori A) and towards the NW (in the direction of $\theta^1$ Ori C).  \\

\noindent 2) The YSO's, 203-504 and 203-506, have a projected separation of only 660\,au, but are extremely different in the nature of their irradiation. 203-504 is a classic bright rimmed proplyd with a teardrop shaped ionisation front due to EUV+FUV irradiation by $\theta^2$\,Ori A. The fact that 203-506 has no ionization boundary indicates that it is not irradiated by the EUV from \tC\ or \tX\ (but is subject to their FUV), a condition satisfied if 203-506 lies closer to the observer than the near ionised layer (NIL). The NIL is about 0.4\,pc in the foreground of $\theta^1$ C \citep{2020ApJ...891...46O} and is of low surface brightness as compared with the MIF. \\

\noindent 3) We detect [Fe\,\textsc{ii}] 8617\AA\ emission from the outflows HH 519 (emanating from 203-504) and HH 520 (emanating from 203-506). We find that the feature previously identified as HH 519 is actually the irradiated rim of the Delta Feature. The new outflow we discover emanating from 203-504, which retains the name HH 519, is a compact outflow tilted at least 45$^{\circ}$ toward the line of sight with red- and blue-shifted velocities of around 130\,km\,s$^{-1}$. The other outflow, HH 520 emanating from 203-506, was previously considered to be monopolar, but we demonstrates that it is in fact bipolar, albeit with one component around a factor 5-10 fainter than the other. \\

\noindent 4) The photoevaporative flow around 203-506 is very bright in [C~{\sc i}] 8727~\AA\ emission. The line is also detected interior to the I-front of 203-504.  These are the first direct spatially resolved detections of atomic carbon emission in external photoevaporative winds. Atomic carbon is predicted to be a good kinematic tracer of the FUV driven component of external winds with ALMA. 203-506 is the best known example to test that prediction, but high spectral resolution is also required. \\

The pair of discs studied in this paper highlights that the surrounding ionisation fronts and shielding can affect the radiation field that discs are exposed to, even when no longer deeply embedded in their natal material, which has been the focus of recent theoretical work \citep{2023MNRAS.520.6159C, 2022MNRAS.512.3788Q, 2023MNRAS.522.1939Q, 2023MNRAS.520.5331W}. Further understanding this interesting phase in disc evolution and environmental impact will require additional studies of discs at the periphery of H\,\textsc{ii} regions.

\section*{Acknowledgements}
We thank Giacomo Beccari, Fuyan Bian and the ESO support team for their help in explaining the instrumental artefacts that appear in the H\,$\alpha$ data. We thank Bo Reipurth for discussions on the identifier for HH 519, and for maintaining the Herbig-Haro catalogue.

TJH acknowledges funding from a Royal Society Dorothy Hodgkin Fellowship and UKRI guaranteed funding for a Horizon Europe ERC consolidator grant (EP/Y024710/1). JRG thanks the Spanish MCINN for funding support under grant PID2019-106110GB-I00.
CFM and MLA are funded by the European Union (ERC, WANDA, 101039452). GB is funded by the European Research Council (ERC) under the European Union’s Horizon 2020 research and innovation programme (Grant agreement No. 853022, PEVAP).
JSK acknowledge support by the National Aeronautics and Space Administration
under Agreement No. 80NSSC21K0593 for the program "Alien Earths".  
Views and opinions expressed are however those of the author(s) only and do not necessarily reflect those of the European Union or the European Research Council Executive Agency. Neither the European Union nor the granting authority can be held responsible for them.

%%%%%%%%%%%%%%%%%%%%%%%%%%%%%%%%%%%%%%%%%%%%%%%%%%
\section*{Data Availability}
The VLT-MUSE data has a 12 month proprietory period, after which the data will be publicly available in the ESO archive. We are happy to share in advance of that date via correspondence. 

\appendix

\vspace{-0.1cm}

\section{Instrumental artefacts}
\label{sec:artefacts}
Our reduced dataset included three small dark artefacts in the H\,\textsc{\,\textsc{ii}} region and near 203-504 in H\,$\alpha$. They were identified as artefacts (rather than tiny globules) because in the individual sub-cubes at different inclination there was a single feature (rather than three) and that was at different locations. We contacted the ESO helpdesk, who also engaged with the team at Paranal, to find an explanation, which is summarised below. 

In NFM, a dichroic reflects infrared light towards an infrared low order sensor, which measures the wave front errors invisible to the laser guide star adaptive optics system. The presence of the dichroic in the light path of MUSE when used in the NFM-AO, has some effect on the quality of the signal. The most noticeable effect of the dichroic is the introduction of a few small regions (spots) of reduced sensitivity, varying from a few up to 10\,per cent compared to the median value. This is likely triggered by the presence of impurities or small particles of dust on the dichroic itself. Using flat fields and images at different rotations efficiently removes  such effects from the combined cubes but sporadically, they can be visible. This is the most likely suggested cause of the instrumental artefacts in our dataset. 

\vspace{-0.1cm}

\section{Explaining the \Ci\ 8727 \AA\ Emission}
\label{sec:appCi}

\Ci\ emission is seen in spectra of the Huygens region in the form of the 
9824.31\,\AA, 9850.34\,\AA\ and 8727.126\,\AA\ lines \citep{2004MNRAS.355..229E}. The longer wavelength lines are forbidden transitions from the 2$\rm s^{2}$ 2$\rm p^{2}$ $\rm ^{1}D_{2}$ level at 1.26 eV above the ground state. The shorter wavelength line is a forbidden transition from the 2$\rm s^{2}$ 2$\rm p^{2}$ $\rm ^{1}S_{0}$ level 2.7 eV above the ground state to the upper level of the longer wavelength lines.

With an ionisation potential of 11.26 eV, atomic carbon is ionised by FUV in much of the PDR.  \cite{1991ApJ...375..630E} argue that recombinations populate the state producing the 8727 \AA\ line, as well as the state giving rise to the 9850\,\AA\ and 9824\,\AA\ lines.  In addition, collisions are also a possible origin of the [C\,\textsc{i}] 8727\,\AA\ emission, however, collisional excitation in the $\sim9000$\,K gas of the Huygens region should lead to  9850\,\AA\ and 9824\,\AA\ lines being much stronger than the [C\,\textsc{i}] 8727\,\AA\ emission, which \cite{2004MNRAS.355..229E} show is not the case. {Specifically,  \cite{2004MNRAS.355..229E} find the 9824\,\AA\ /~8727\,\AA\ flux ratio is of order unity and the 9850\,\AA\ / 8727\,\AA\ flux ratio is $\sim 5$, which is consistent with recombination excitation in a PDR \citep{1991ApJ...375..630E}.} We hence conclude that the most likely source of the 8727\,\AA\  emission is due to photoionizations by FUV followed by recombinations rather than collisional excitation.

\vspace{-0.1cm}

\section{Explaining the \oi\ 8446\,\AA\ Emission}
\label{sec:appveldif} 

The surprising presence of \oi\ 8446 \AA\ lines in the spectra of gaseous nebulae was first explained as the result of fluorescence by \Lyb\ by \citet{1928ApJ....67....1B}, with the details of the process being explained more recently in Section 4.7 and Figure 4.7 of \cite{2006agna.book.....O}. \Lyb\ (1025.72 \AA ) has a near coincidence in wavelength with a ground state line of O\,\textsc{i} at 1025.76 \AA , driving excitation of the upper state of the neutral oxygen, followed by spontaneous decay. Production of the line requires both the presence of neutral oxygen and \Lyb\ photons from ionized hydrogen. The \Lyb\ photons could come from co-existing ionized hydrogen interacting with neutral oxygen produced by the mechanism of charge exchange, where ionized oxygen gain an electron from the highly ionized hydrogen gas. An alternative source of \Lyb\ is emission from the ionized hydrogen zone that begins at the ionization boundary.

The ionization balance in the Huygens Region is very dependent on the distance from \tC, as shown in the multi-sample study of \citet{2010AJ....140..985O}. Because of problems of correcting for
night-sky \oi\ 6300 \AA\, the pattern of appearance of \oi\ could not be traced, but the pattern of increasing relative strength of low ionization lines is shown in their Figure 11 (\sii) and Figure 12 (\nii). This pattern of increasing relative strength of \sii\ and \nii\ leads one to expect an increasing relative strength of \oi , which explains the larger relative strengths of \oi\ 8446 \AA\ in the outer regions in Table~\ref{tab:fluxes}.

\begin{figure}
    \centering
    \includegraphics[width=\columnwidth]{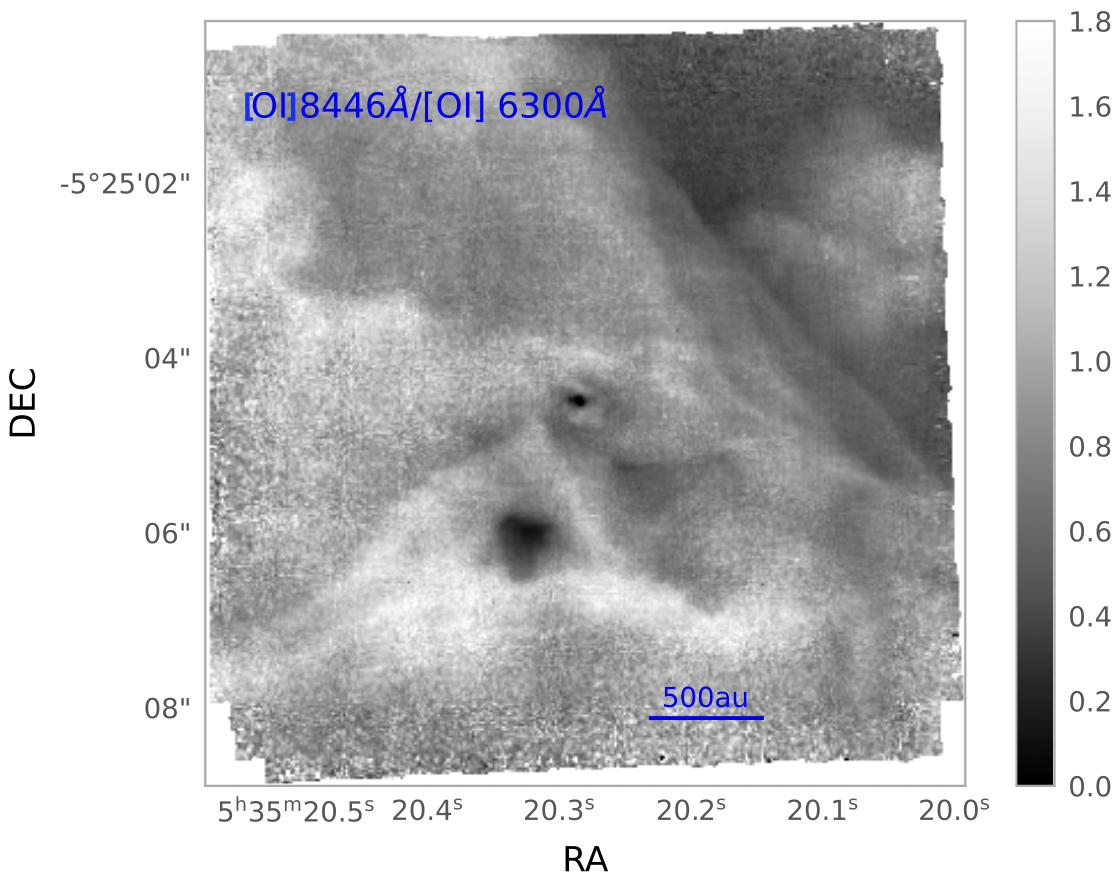}
    \caption{The ratio of [O\,\textsc{i}]\ 8446\,\AA\ to [O\,\textsc{i}]\ 6300\,\AA.The strength of [O\,\textsc{i}] 8446\,\AA\ is driven by resonance with Lyman\,$\beta$ , c. f. 
    \citet{2006agna.book.....O} and that of [O\,\textsc{i}]\ 6300\, \AA\ by collisional excitation (except within the external photoevaporative winds) as explained in section \ref{sec:OI}. The variations are largely caused by velocity differences of the emitter and absorber ions. }
    \label{fig:OI8446on6300}
\end{figure}

The relative velocities of the \Lyb\ emitting (hydrogen) and \Lyb\ absorbing (oxygen) gas plays a role in \oi\ 8446 \AA\ emission.If both oxygen and hydrogen have the same average velocity, the peak of the oxygen line profile is 11.7 \kms\ to the red from the peak of the \Lyb\ line.At  9000 K, the full width at half maximum intensity is 20.4 \kms\ for hydrogen and 5.1 \kms\ for oxygen. The imperfect alignment of the two lines (the oxygen line  peak is about at the half-power point of the hydrogen profile) means that the efficiency of the production of \oi\ is sensitive to their relative velocities. For example, shifting the oxygen to a lower velocity difference would increase the strength of the 8446 \AA\ line.  Table 5 of \citet{2020ApJ...891...46O} summarizes the velocities of difference ions in the MIF in the central Huygens Region, giving 27$\pm$2 \kms for \oi\ and 17$\pm$2 \kms\ for hydrogen Balmer lines. This means that if the \Lyb\ photons pumping \oi\ 8446 \AA\ arise from a different zone, the separation of the peaks of the two lines has increased by 10 \kms\ and the enhancement of \oi\ 8446 \AA\ would be reduced, as compared with a condition where the \Lyb\ source and the neutral oxygen have the same velocity. 

The velocity difference in the central Huygens Region is due to photoevaporative flow away from the hydrogen ionization front that is being viewed face-on. If the flow velocity is less in this outer region where the EUV is weaker, then the \oi\ 8446 \AA\ emissivity would be increased, accounting for the enhancement seen in Table~\ref{tab:fluxes}.

Differences of relative velocities are likely to play a role within the Delta Feature. In Figure \ref{fig:OI8446on6300} we show the ratio of the [O\,\textsc{i}] 8446\AA\ to the [O\,\textsc{i}] 6300\AA\ line. Ignoring the photoevaporating discs (where processes such as OH dissociation determine the 6300\,\AA\ line flux) in the surrounding medium we see the results of differences of velocity of the emitting and absorbing gas, with the more intense [O\,\textsc{i}] 8446 \AA\ arising from greater overlap of the \Lyb\ and neutral oxygen line profiles while the [O\,\textsc{i}]\ 6300\ emission arises from collisional excitation by the electron gas, independent of the gas velocity. At the well defined NE and NW boundaries of the Delta Feature abrupt changes of relative velocity are to be expected, as shown in high velocity resolution studies of the Bar \citep{2017MNRAS.464.4835O}, if this feature is a discrete physical structure.  The slowly changing values of the ratio along the south boundary of the Delta Feature are consistent with the idea that the boundary there is more gradual.



\bibliographystyle{mnras}
\bibliography{bibfile.bib} % if your bibtex file is called example.bib

\begin{thebibliography}{}
\makeatletter
\relax
\def\mn@urlcharsother{\let\do\@makeother \do\$\do\&\do\#\do\^\do\_\do\%\do\~}
\def\mn@doi{\begingroup\mn@urlcharsother \@ifnextchar [ {\mn@doi@}
  {\mn@doi@[]}}
\def\mn@doi@[#1]#2{\def\@tempa{#1}\ifx\@tempa\@empty \href
  {http://dx.doi.org/#2} {doi:#2}\else \href {http://dx.doi.org/#2} {#1}\fi
  \endgroup}
\def\mn@eprint#1#2{\mn@eprint@#1:#2::\@nil}
\def\mn@eprint@arXiv#1{\href {http://arxiv.org/abs/#1} {{\tt arXiv:#1}}}
\def\mn@eprint@dblp#1{\href {http://dblp.uni-trier.de/rec/bibtex/#1.xml}
  {dblp:#1}}
\def\mn@eprint@#1:#2:#3:#4\@nil{\def\@tempa {#1}\def\@tempb {#2}\def\@tempc
  {#3}\ifx \@tempc \@empty \let \@tempc \@tempb \let \@tempb \@tempa \fi \ifx
  \@tempb \@empty \def\@tempb {arXiv}\fi \@ifundefined
  {mn@eprint@\@tempb}{\@tempb:\@tempc}{\expandafter \expandafter \csname
  mn@eprint@\@tempb\endcsname \expandafter{\@tempc}}}

\bibitem[\protect\citeauthoryear{{ALMA Partnership} et~al.,}{{ALMA Partnership}
  et~al.}{2015}]{2015ApJ...808L...3A}
{ALMA Partnership} et~al., 2015, \mn@doi [\apjl] {10.1088/2041-8205/808/1/L3},
  \href {https://ui.adsabs.harvard.edu/abs/2015ApJ...808L...3A} {808, L3}

\bibitem[\protect\citeauthoryear{{Abel}, {Ferland}  \& {O'Dell}}{{Abel}
  et~al.}{2019}]{2019ApJ...881..130A}
{Abel} N.~P.,  {Ferland} G.~J.,   {O'Dell} C.~R.,  2019, \mn@doi [\apj]
  {10.3847/1538-4357/ab2a6e}, \href
  {https://ui.adsabs.harvard.edu/abs/2019ApJ...881..130A} {881, 130}

\bibitem[\protect\citeauthoryear{{Abt}, {Wang}  \& {Cardona}}{{Abt}
  et~al.}{1991}]{1991ApJ...367..155A}
{Abt} H.~A.,  {Wang} R.,   {Cardona} O.,  1991, \mn@doi [\apj]
  {10.1086/169611}, \href
  {https://ui.adsabs.harvard.edu/abs/1991ApJ...367..155A} {367, 155}

\bibitem[\protect\citeauthoryear{{Adams}, {Hollenbach}, {Laughlin}  \&
  {Gorti}}{{Adams} et~al.}{2004}]{2004ApJ...611..360A}
{Adams} F.~C.,  {Hollenbach} D.,  {Laughlin} G.,   {Gorti} U.,  2004, \mn@doi
  [\apj] {10.1086/421989}, \href
  {https://ui.adsabs.harvard.edu/abs/2004ApJ...611..360A} {611, 360}

\bibitem[\protect\citeauthoryear{{Agra-Amboage}, {Dougados}, {Cabrit}  \&
  {Reunanen}}{{Agra-Amboage} et~al.}{2011}]{2011A&A...532A..59A}
{Agra-Amboage} V.,  {Dougados} C.,  {Cabrit} S.,   {Reunanen} J.,  2011,
  \mn@doi [\aap] {10.1051/0004-6361/201015886}, \href
  {https://ui.adsabs.harvard.edu/abs/2011A&A...532A..59A} {532, A59}

\bibitem[\protect\citeauthoryear{{Andree-Labsch}, {Ossenkopf-Okada}  \&
  {R{\"o}llig}}{{Andree-Labsch} et~al.}{2017}]{2017A&A...598A...2A}
{Andree-Labsch} S.,  {Ossenkopf-Okada} V.,   {R{\"o}llig} M.,  2017, \mn@doi
  [\aap] {10.1051/0004-6361/201424287}, \href
  {https://ui.adsabs.harvard.edu/abs/2017A&A...598A...2A} {598, A2}

\bibitem[\protect\citeauthoryear{{Andrews} et~al.,}{{Andrews}
  et~al.}{2018}]{2018ApJ...869L..41A}
{Andrews} S.~M.,  et~al., 2018, \mn@doi [\apjl] {10.3847/2041-8213/aaf741},
  \href {https://ui.adsabs.harvard.edu/abs/2018ApJ...869L..41A} {869, L41}

\bibitem[\protect\citeauthoryear{{Bacon} et~al.,}{{Bacon}
  et~al.}{2010}]{2010SPIE.7735E..08B}
{Bacon} R.,  et~al., 2010, in {McLean} I.~S.,  {Ramsay} S.~K.,   {Takami} H.,
  eds,  Society of Photo-Optical Instrumentation Engineers (SPIE) Conference
  Series Vol. 7735, Ground-based and Airborne Instrumentation for Astronomy
  III. p. 773508 (\mn@eprint {arXiv} {2211.16795}), \mn@doi{10.1117/12.856027}

\bibitem[\protect\citeauthoryear{{Baldwin}, {Ferland}, {Martin}, {Corbin},
  {Cota}, {Peterson}  \& {Slettebak}}{{Baldwin}
  et~al.}{1991}]{1991ApJ...374..580B}
{Baldwin} J.~A.,  {Ferland} G.~J.,  {Martin} P.~G.,  {Corbin} M.~R.,  {Cota}
  S.~A.,  {Peterson} B.~M.,   {Slettebak} A.,  1991, \mn@doi [\apj]
  {10.1086/170146}, \href
  {https://ui.adsabs.harvard.edu/abs/1991ApJ...374..580B} {374, 580}

\bibitem[\protect\citeauthoryear{{Ballabio}, {Haworth}  \& {Henney}}{{Ballabio}
  et~al.}{2023}]{2023MNRAS.518.5563B}
{Ballabio} G.,  {Haworth} T.~J.,   {Henney} W.~J.,  2023, \mn@doi [\mnras]
  {10.1093/mnras/stac3467}, \href
  {https://ui.adsabs.harvard.edu/abs/2023MNRAS.518.5563B} {518, 5563}

\bibitem[\protect\citeauthoryear{{Bally}, {O'Dell}  \& {McCaughrean}}{{Bally}
  et~al.}{2000}]{2000AJ....119.2919B}
{Bally} J.,  {O'Dell} C.~R.,   {McCaughrean} M.~J.,  2000, \mn@doi [\aj]
  {10.1086/301385}, \href
  {https://ui.adsabs.harvard.edu/abs/2000AJ....119.2919B} {119, 2919}

\bibitem[\protect\citeauthoryear{{Bern{\'e}} et~al.,}{{Bern{\'e}}
  et~al.}{2022}]{2022PASP..134e4301B}
{Bern{\'e}} O.,  et~al., 2022, \mn@doi [\pasp] {10.1088/1538-3873/ac604c},
  \href {https://ui.adsabs.harvard.edu/abs/2022PASP..134e4301B} {134, 054301}

\bibitem[\protect\citeauthoryear{{Blagrave}, {Martin}, {Rubin}, {Dufour},
  {Baldwin}, {Hester}  \& {Walter}}{{Blagrave}
  et~al.}{2007}]{2007ApJ...655..299B}
{Blagrave} K.~P.~M.,  {Martin} P.~G.,  {Rubin} R.~H.,  {Dufour} R.~J.,
  {Baldwin} J.~A.,  {Hester} J.~J.,   {Walter} D.~K.,  2007, \mn@doi [\apj]
  {10.1086/510151}, \href
  {https://ui.adsabs.harvard.edu/abs/2007ApJ...655..299B} {655, 299}

\bibitem[\protect\citeauthoryear{{Bowen}}{{Bowen}}{1928}]{1928ApJ....67....1B}
{Bowen} I.~S.,  1928, \mn@doi [\apj] {10.1086/143091}, \href
  {https://ui.adsabs.harvard.edu/abs/1928ApJ....67....1B} {67, 1}

\bibitem[\protect\citeauthoryear{{Boyden} \& {Eisner}}{{Boyden} \&
  {Eisner}}{2020}]{2020ApJ...894...74B}
{Boyden} R.~D.,  {Eisner} J.~A.,  2020, \mn@doi [\apj]
  {10.3847/1538-4357/ab86b7}, \href
  {https://ui.adsabs.harvard.edu/abs/2020ApJ...894...74B} {894, 74}

\bibitem[\protect\citeauthoryear{{Boyden} \& {Eisner}}{{Boyden} \&
  {Eisner}}{2023}]{2023ApJ...947....7B}
{Boyden} R.~D.,  {Eisner} J.~A.,  2023, \mn@doi [\apj]
  {10.3847/1538-4357/acaf77}, \href
  {https://ui.adsabs.harvard.edu/abs/2023ApJ...947....7B} {947, 7}

\bibitem[\protect\citeauthoryear{{Bregman}, {Allamandola}, {Tielens}, {Geballe}
   \& {Witteborn}}{{Bregman} et~al.}{1989}]{1989ApJ...344..791B}
{Bregman} J.~D.,  {Allamandola} L.~J.,  {Tielens} A.~G.~G.~M.,  {Geballe}
  T.~R.,   {Witteborn} F.~C.,  1989, \mn@doi [\apj] {10.1086/167844}, \href
  {https://ui.adsabs.harvard.edu/abs/1989ApJ...344..791B} {344, 791}

\bibitem[\protect\citeauthoryear{{Cardelli}, {Clayton}  \& {Mathis}}{{Cardelli}
  et~al.}{1989}]{1989ApJ...345..245C}
{Cardelli} J.~A.,  {Clayton} G.~C.,   {Mathis} J.~S.,  1989, \mn@doi [\apj]
  {10.1086/167900}, \href
  {https://ui.adsabs.harvard.edu/abs/1989ApJ...345..245C} {345, 245}

\bibitem[\protect\citeauthoryear{{Champion}, {Bern{\'e}}, {Vicente}, {Kamp},
  {Le Petit}, {Gusdorf}, {Joblin}  \& {Goicoechea}}{{Champion}
  et~al.}{2017}]{2017A&A...604A..69C}
{Champion} J.,  {Bern{\'e}} O.,  {Vicente} S.,  {Kamp} I.,  {Le Petit} F.,
  {Gusdorf} A.,  {Joblin} C.,   {Goicoechea} J.~R.,  2017, \mn@doi [\aap]
  {10.1051/0004-6361/201629404}, \href
  {https://ui.adsabs.harvard.edu/abs/2017A&A...604A..69C} {604, A69}

\bibitem[\protect\citeauthoryear{{Chandar}}{{Chandar}}{2009}]{2009Ap&SS.324..315C}
{Chandar} R.,  2009, \mn@doi [\apss] {10.1007/s10509-009-0105-8}, \href
  {https://ui.adsabs.harvard.edu/abs/2009Ap&SS.324..315C} {324, 315}

\bibitem[\protect\citeauthoryear{{Clarke} \& {Owen}}{{Clarke} \&
  {Owen}}{2015}]{2015MNRAS.446.2944C}
{Clarke} C.~J.,  {Owen} J.~E.,  2015, \mn@doi [\mnras] {10.1093/mnras/stu2248},
  \href {https://ui.adsabs.harvard.edu/abs/2015MNRAS.446.2944C} {446, 2944}

\bibitem[\protect\citeauthoryear{{Concha-Ram{\'\i}rez}, {Wilhelm}  \&
  {Portegies Zwart}}{{Concha-Ram{\'\i}rez} et~al.}{2023}]{2023MNRAS.520.6159C}
{Concha-Ram{\'\i}rez} F.,  {Wilhelm} M. J.~C.,   {Portegies Zwart} S.,  2023,
  \mn@doi [\mnras] {10.1093/mnras/stac1733}, \href
  {https://ui.adsabs.harvard.edu/abs/2023MNRAS.520.6159C} {520, 6159}

\bibitem[\protect\citeauthoryear{{Cuadrado}, {Goicoechea}, {Pilleri},
  {Cernicharo}, {Fuente}  \& {Joblin}}{{Cuadrado}
  et~al.}{2015}]{2015A&A...575A..82C}
{Cuadrado} S.,  {Goicoechea} J.~R.,  {Pilleri} P.,  {Cernicharo} J.,  {Fuente}
  A.,   {Joblin} C.,  2015, \mn@doi [\aap] {10.1051/0004-6361/201424568}, \href
  {https://ui.adsabs.harvard.edu/abs/2015A&A...575A..82C} {575, A82}

\bibitem[\protect\citeauthoryear{{Cuello} et~al.,}{{Cuello}
  et~al.}{2019}]{2019MNRAS.483.4114C}
{Cuello} N.,  et~al., 2019, \mn@doi [\mnras] {10.1093/mnras/sty3325}, \href
  {https://ui.adsabs.harvard.edu/abs/2019MNRAS.483.4114C} {483, 4114}

\bibitem[\protect\citeauthoryear{{Cuello} et~al.,}{{Cuello}
  et~al.}{2020}]{2020MNRAS.491..504C}
{Cuello} N.,  et~al., 2020, \mn@doi [\mnras] {10.1093/mnras/stz2938}, \href
  {https://ui.adsabs.harvard.edu/abs/2020MNRAS.491..504C} {491, 504}

\bibitem[\protect\citeauthoryear{{Cuello}, {M{\'e}nard}  \& {Price}}{{Cuello}
  et~al.}{2023}]{2023EPJP..138...11C}
{Cuello} N.,  {M{\'e}nard} F.,   {Price} D.~J.,  2023, \mn@doi [European
  Physical Journal Plus] {10.1140/epjp/s13360-022-03602-w}, \href
  {https://ui.adsabs.harvard.edu/abs/2023EPJP..138...11C} {138, 11}

\bibitem[\protect\citeauthoryear{{Dale} \& {Bonnell}}{{Dale} \&
  {Bonnell}}{2011}]{2011MNRAS.414..321D}
{Dale} J.~E.,  {Bonnell} I.,  2011, \mn@doi [\mnras]
  {10.1111/j.1365-2966.2011.18392.x}, \href
  {https://ui.adsabs.harvard.edu/abs/2011MNRAS.414..321D} {414, 321}

\bibitem[\protect\citeauthoryear{{Dale}, {Ercolano}  \& {Bonnell}}{{Dale}
  et~al.}{2012}]{2012MNRAS.424..377D}
{Dale} J.~E.,  {Ercolano} B.,   {Bonnell} I.~A.,  2012, \mn@doi [\mnras]
  {10.1111/j.1365-2966.2012.21205.x}, \href
  {https://ui.adsabs.harvard.edu/abs/2012MNRAS.424..377D} {424, 377}

\bibitem[\protect\citeauthoryear{{Dale}, {Ngoumou}, {Ercolano}  \&
  {Bonnell}}{{Dale} et~al.}{2014}]{2014MNRAS.442..694D}
{Dale} J.~E.,  {Ngoumou} J.,  {Ercolano} B.,   {Bonnell} I.~A.,  2014, \mn@doi
  [\mnras] {10.1093/mnras/stu816}, \href
  {https://ui.adsabs.harvard.edu/abs/2014MNRAS.442..694D} {442, 694}

\bibitem[\protect\citeauthoryear{{Dipierro}, {Price}, {Laibe}, {Hirsh},
  {Cerioli}  \& {Lodato}}{{Dipierro} et~al.}{2015}]{2015MNRAS.453L..73D}
{Dipierro} G.,  {Price} D.,  {Laibe} G.,  {Hirsh} K.,  {Cerioli} A.,   {Lodato}
  G.,  2015, \mn@doi [\mnras] {10.1093/mnrasl/slv105}, \href
  {https://ui.adsabs.harvard.edu/abs/2015MNRAS.453L..73D} {453, L73}

\bibitem[\protect\citeauthoryear{{Eisner} et~al.,}{{Eisner}
  et~al.}{2018}]{2018ApJ...860...77E}
{Eisner} J.~A.,  et~al., 2018, \mn@doi [\apj] {10.3847/1538-4357/aac3e2}, \href
  {https://ui.adsabs.harvard.edu/abs/2018ApJ...860...77E} {860, 77}

\bibitem[\protect\citeauthoryear{{Escalante}, {Sternberg}  \&
  {Dalgarno}}{{Escalante} et~al.}{1991}]{1991ApJ...375..630E}
{Escalante} V.,  {Sternberg} A.,   {Dalgarno} A.,  1991, \mn@doi [\apj]
  {10.1086/170225}, \href
  {https://ui.adsabs.harvard.edu/abs/1991ApJ...375..630E} {375, 630}

\bibitem[\protect\citeauthoryear{{Esteban}, {Peimbert}, {Garc{\'\i}a-Rojas},
  {Ruiz}, {Peimbert}  \& {Rodr{\'\i}guez}}{{Esteban}
  et~al.}{2004}]{2004MNRAS.355..229E}
{Esteban} C.,  {Peimbert} M.,  {Garc{\'\i}a-Rojas} J.,  {Ruiz} M.~T.,
  {Peimbert} A.,   {Rodr{\'\i}guez} M.,  2004, \mn@doi [\mnras]
  {10.1111/j.1365-2966.2004.08313.x}, \href
  {https://ui.adsabs.harvard.edu/abs/2004MNRAS.355..229E} {355, 229}

\bibitem[\protect\citeauthoryear{{Facchini}, {Clarke}  \& {Bisbas}}{{Facchini}
  et~al.}{2016}]{2016MNRAS.457.3593F}
{Facchini} S.,  {Clarke} C.~J.,   {Bisbas} T.~G.,  2016, \mn@doi [\mnras]
  {10.1093/mnras/stw240}, \href
  {https://ui.adsabs.harvard.edu/abs/2016MNRAS.457.3593F} {457, 3593}

\bibitem[\protect\citeauthoryear{{Fall}, {Krumholz}  \& {Matzner}}{{Fall}
  et~al.}{2010}]{2010ApJ...710L.142F}
{Fall} S.~M.,  {Krumholz} M.~R.,   {Matzner} C.~D.,  2010, \mn@doi [\apjl]
  {10.1088/2041-8205/710/2/L142}, \href
  {https://ui.adsabs.harvard.edu/abs/2010ApJ...710L.142F} {710, L142}

\bibitem[\protect\citeauthoryear{{Fang}, {Kim}, {Pascucci}, {Apai}  \&
  {Manara}}{{Fang} et~al.}{2016}]{2016ApJ...833L..16F}
{Fang} M.,  {Kim} J.~S.,  {Pascucci} I.,  {Apai} D.,   {Manara} C.~F.,  2016,
  \mn@doi [\apjl] {10.3847/2041-8213/833/2/L16}, \href
  {https://ui.adsabs.harvard.edu/abs/2016ApJ...833L..16F} {833, L16}

\bibitem[\protect\citeauthoryear{{Fang}, {Kim}, {Pascucci}  \& {Apai}}{{Fang}
  et~al.}{2021}]{2021ApJ...908...49F}
{Fang} M.,  {Kim} J.~S.,  {Pascucci} I.,   {Apai} D.,  2021, \mn@doi [\apj]
  {10.3847/1538-4357/abcec8}, \href
  {https://ui.adsabs.harvard.edu/abs/2021ApJ...908...49F} {908, 49}

\bibitem[\protect\citeauthoryear{{Ferland}, {Korista}, {Verner}, {Ferguson},
  {Kingdon}  \& {Verner}}{{Ferland} et~al.}{1998}]{1998PASP..110..761F}
{Ferland} G.~J.,  {Korista} K.~T.,  {Verner} D.~A.,  {Ferguson} J.~W.,
  {Kingdon} J.~B.,   {Verner} E.~M.,  1998, \mn@doi [\pasp] {10.1086/316190},
  \href {https://ui.adsabs.harvard.edu/abs/1998PASP..110..761F} {110, 761}

\bibitem[\protect\citeauthoryear{{Ferland} et~al.,}{{Ferland}
  et~al.}{2013}]{2013RMxAA..49..137F}
{Ferland} G.~J.,  et~al., 2013, \mn@doi [\rmxaa] {10.48550/arXiv.1302.4485},
  \href {https://ui.adsabs.harvard.edu/abs/2013RMxAA..49..137F} {49, 137}

\bibitem[\protect\citeauthoryear{{Ferland} et~al.,}{{Ferland}
  et~al.}{2017}]{2017RMxAA..53..385F}
{Ferland} G.~J.,  et~al., 2017, \mn@doi [\rmxaa] {10.48550/arXiv.1705.10877},
  \href {https://ui.adsabs.harvard.edu/abs/2017RMxAA..53..385F} {53, 385}

\bibitem[\protect\citeauthoryear{{Freudling}, {Romaniello}, {Bramich},
  {Ballester}, {Forchi}, {Garc{\'\i}a-Dabl{\'o}}, {Moehler}  \&
  {Neeser}}{{Freudling} et~al.}{2013}]{2013A&A...559A..96F}
{Freudling} W.,  {Romaniello} M.,  {Bramich} D.~M.,  {Ballester} P.,  {Forchi}
  V.,  {Garc{\'\i}a-Dabl{\'o}} C.~E.,  {Moehler} S.,   {Neeser} M.~J.,  2013,
  \mn@doi [\aap] {10.1051/0004-6361/201322494}, \href
  {https://ui.adsabs.harvard.edu/abs/2013A&A...559A..96F} {559, A96}

\bibitem[\protect\citeauthoryear{{Fukui}, {Habe}, {Inoue}, {Enokiya}  \&
  {Tachihara}}{{Fukui} et~al.}{2021}]{2021PASJ...73S...1F}
{Fukui} Y.,  {Habe} A.,  {Inoue} T.,  {Enokiya} R.,   {Tachihara} K.,  2021,
  \mn@doi [\pasj] {10.1093/pasj/psaa103}, \href
  {https://ui.adsabs.harvard.edu/abs/2021PASJ...73S...1F} {73, S1}

\bibitem[\protect\citeauthoryear{{Gaia Collaboration} et~al.,}{{Gaia
  Collaboration} et~al.}{2016}]{2016A&A...595A...1G}
{Gaia Collaboration} et~al., 2016, \mn@doi [\aap]
  {10.1051/0004-6361/201629272}, \href
  {https://ui.adsabs.harvard.edu/abs/2016A&A...595A...1G} {595, A1}

\bibitem[\protect\citeauthoryear{{Gaia Collaboration} et~al.,}{{Gaia
  Collaboration} et~al.}{2018}]{2018A&A...616A...1G}
{Gaia Collaboration} et~al., 2018, \mn@doi [\aap]
  {10.1051/0004-6361/201833051}, \href
  {https://ui.adsabs.harvard.edu/abs/2018A&A...616A...1G} {616, A1}

\bibitem[\protect\citeauthoryear{{Gaia Collaboration} et~al.,}{{Gaia
  Collaboration} et~al.}{2022}]{2022arXiv220800211G}
{Gaia Collaboration} et~al., 2022, \mn@doi [arXiv e-prints]
  {10.48550/arXiv.2208.00211}, \href
  {https://ui.adsabs.harvard.edu/abs/2022arXiv220800211G} {p. arXiv:2208.00211}

\bibitem[\protect\citeauthoryear{{Garc{\'\i}a-D{\'\i}az} \&
  {Henney}}{{Garc{\'\i}a-D{\'\i}az} \& {Henney}}{2007}]{2007AJ....133..952G}
{Garc{\'\i}a-D{\'\i}az} M.~T.,  {Henney} W.~J.,  2007, \mn@doi [\aj]
  {10.1086/510621}, \href
  {https://ui.adsabs.harvard.edu/abs/2007AJ....133..952G} {133, 952}

\bibitem[\protect\citeauthoryear{{Garufi} et~al.,}{{Garufi}
  et~al.}{2022}]{2022A&A...658A.104G}
{Garufi} A.,  et~al., 2022, \mn@doi [\aap] {10.1051/0004-6361/202141264}, \href
  {https://ui.adsabs.harvard.edu/abs/2022A&A...658A.104G} {658, A104}

\bibitem[\protect\citeauthoryear{{Gaudi}, {Meyer}  \& {Christiansen}}{{Gaudi}
  et~al.}{2021}]{2021exbi.book....2G}
{Gaudi} B.~S.,  {Meyer} M.,   {Christiansen} J.,  2021, in {Madhusudhan} N.,
  ed., , ExoFrontiers; Big Questions in Exoplanetary Science.
pp~2--1, \mn@doi{10.1088/2514-3433/abfa8fch2}

\bibitem[\protect\citeauthoryear{{Giannini} et~al.,}{{Giannini}
  et~al.}{2013}]{2013ApJ...778...71G}
{Giannini} T.,  et~al., 2013, \mn@doi [\apj] {10.1088/0004-637X/778/1/71},
  \href {https://ui.adsabs.harvard.edu/abs/2013ApJ...778...71G} {778, 71}

\bibitem[\protect\citeauthoryear{{Ginski} et~al.,}{{Ginski}
  et~al.}{2021}]{2021ApJ...908L..25G}
{Ginski} C.,  et~al., 2021, \mn@doi [\apjl] {10.3847/2041-8213/abdf57}, \href
  {https://ui.adsabs.harvard.edu/abs/2021ApJ...908L..25G} {908, L25}

\bibitem[\protect\citeauthoryear{{Goicoechea} et~al.,}{{Goicoechea}
  et~al.}{2016}]{2016Natur.537..207G}
{Goicoechea} J.~R.,  et~al., 2016, \mn@doi [\nat] {10.1038/nature18957}, \href
  {https://ui.adsabs.harvard.edu/abs/2016Natur.537..207G} {537, 207}

\bibitem[\protect\citeauthoryear{{Greve}, {Castles}  \& {McKeith}}{{Greve}
  et~al.}{1994}]{1994A&A...284..919G}
{Greve} A.,  {Castles} J.,   {McKeith} C.~D.,  1994, \aap, \href
  {https://ui.adsabs.harvard.edu/abs/1994A&A...284..919G} {284, 919}

\bibitem[\protect\citeauthoryear{{Gritschneder}, {Naab}, {Walch}, {Burkert}  \&
  {Heitsch}}{{Gritschneder} et~al.}{2009}]{2009ApJ...694L..26G}
{Gritschneder} M.,  {Naab} T.,  {Walch} S.,  {Burkert} A.,   {Heitsch} F.,
  2009, \mn@doi [\apjl] {10.1088/0004-637X/694/1/L26}, \href
  {https://ui.adsabs.harvard.edu/abs/2009ApJ...694L..26G} {694, L26}

\bibitem[\protect\citeauthoryear{{Gritschneder}, {Burkert}, {Naab}  \&
  {Walch}}{{Gritschneder} et~al.}{2010}]{2010ApJ...723..971G}
{Gritschneder} M.,  {Burkert} A.,  {Naab} T.,   {Walch} S.,  2010, \mn@doi
  [\apj] {10.1088/0004-637X/723/2/971}, \href
  {https://ui.adsabs.harvard.edu/abs/2010ApJ...723..971G} {723, 971}

\bibitem[\protect\citeauthoryear{{Gro{\ss}schedl} et~al.,}{{Gro{\ss}schedl}
  et~al.}{2018}]{2018A&A...619A.106G}
{Gro{\ss}schedl} J.~E.,  et~al., 2018, \mn@doi [\aap]
  {10.1051/0004-6361/201833901}, \href
  {https://ui.adsabs.harvard.edu/abs/2018A&A...619A.106G} {619, A106}

\bibitem[\protect\citeauthoryear{{Grudi{\'c}}, {Guszejnov}, {Hopkins}, {Offner}
   \& {Faucher-Gigu{\`e}re}}{{Grudi{\'c}} et~al.}{2021}]{2021MNRAS.506.2199G}
{Grudi{\'c}} M.~Y.,  {Guszejnov} D.,  {Hopkins} P.~F.,  {Offner} S. S.~R.,
  {Faucher-Gigu{\`e}re} C.-A.,  2021, \mn@doi [\mnras]
  {10.1093/mnras/stab1347}, \href
  {https://ui.adsabs.harvard.edu/abs/2021MNRAS.506.2199G} {506, 2199}

\bibitem[\protect\citeauthoryear{{Grudi{\'c}}, {Guszejnov}, {Offner}, {Rosen},
  {Raju}, {Faucher-Gigu{\`e}re}  \& {Hopkins}}{{Grudi{\'c}}
  et~al.}{2022}]{2022MNRAS.512..216G}
{Grudi{\'c}} M.~Y.,  {Guszejnov} D.,  {Offner} S. S.~R.,  {Rosen} A.~L.,
  {Raju} A.~N.,  {Faucher-Gigu{\`e}re} C.-A.,   {Hopkins} P.~F.,  2022, \mn@doi
  [\mnras] {10.1093/mnras/stac526}, \href
  {https://ui.adsabs.harvard.edu/abs/2022MNRAS.512..216G} {512, 216}

\bibitem[\protect\citeauthoryear{{Gupta} et~al.,}{{Gupta}
  et~al.}{2023}]{2023A&A...670L...8G}
{Gupta} A.,  et~al., 2023, \mn@doi [\aap] {10.1051/0004-6361/202245254}, \href
  {https://ui.adsabs.harvard.edu/abs/2023A&A...670L...8G} {670, L8}

\bibitem[\protect\citeauthoryear{{Guszejnov}, {Grudi{\'c}}, {Offner},
  {Faucher-Gigu{\`e}re}, {Hopkins}  \& {Rosen}}{{Guszejnov}
  et~al.}{2022}]{2022MNRAS.515.4929G}
{Guszejnov} D.,  {Grudi{\'c}} M.~Y.,  {Offner} S. S.~R.,  {Faucher-Gigu{\`e}re}
  C.-A.,  {Hopkins} P.~F.,   {Rosen} A.~L.,  2022, \mn@doi [\mnras]
  {10.1093/mnras/stac2060}, \href
  {https://ui.adsabs.harvard.edu/abs/2022MNRAS.515.4929G} {515, 4929}

\bibitem[\protect\citeauthoryear{{Habart} et~al.,}{{Habart}
  et~al.}{2023}]{2023A&A...673A.149H}
{Habart} E.,  et~al., 2023, \mn@doi [\aap] {10.1051/0004-6361/202244034}, \href
  {https://ui.adsabs.harvard.edu/abs/2023A&A...673A.149H} {673, A149}

\bibitem[\protect\citeauthoryear{{Haffert}, {Bohn}, {de Boer}, {Snellen},
  {Brinchmann}, {Girard}, {Keller}  \& {Bacon}}{{Haffert}
  et~al.}{2019}]{2019NatAs...3..749H}
{Haffert} S.~Y.,  {Bohn} A.~J.,  {de Boer} J.,  {Snellen} I.~A.~G.,
  {Brinchmann} J.,  {Girard} J.~H.,  {Keller} C.~U.,   {Bacon} R.,  2019,
  \mn@doi [Nature Astronomy] {10.1038/s41550-019-0780-5}, \href
  {https://ui.adsabs.harvard.edu/abs/2019NatAs...3..749H} {3, 749}

\bibitem[\protect\citeauthoryear{{Hartigan} \& {Morse}}{{Hartigan} \&
  {Morse}}{2007}]{2007ApJ...660..426H}
{Hartigan} P.,  {Morse} J.,  2007, \mn@doi [\apj] {10.1086/513015}, \href
  {https://ui.adsabs.harvard.edu/abs/2007ApJ...660..426H} {660, 426}

\bibitem[\protect\citeauthoryear{{Hartigan}, {Morse}, {Reipurth}, {Heathcote}
  \& {Bally}}{{Hartigan} et~al.}{2001}]{2001ApJ...559L.157H}
{Hartigan} P.,  {Morse} J.~A.,  {Reipurth} B.,  {Heathcote} S.,   {Bally} J.,
  2001, \mn@doi [\apjl] {10.1086/323976}, \href
  {https://ui.adsabs.harvard.edu/abs/2001ApJ...559L.157H} {559, L157}

\bibitem[\protect\citeauthoryear{{Hartigan}, {Raymond}  \&
  {Pierson}}{{Hartigan} et~al.}{2004}]{2004ApJ...614L..69H}
{Hartigan} P.,  {Raymond} J.,   {Pierson} R.,  2004, \mn@doi [\apjl]
  {10.1086/425322}, \href
  {https://ui.adsabs.harvard.edu/abs/2004ApJ...614L..69H} {614, L69}

\bibitem[\protect\citeauthoryear{{Hartigan}, {Heathcote}, {Morse}, {Reipurth}
  \& {Bally}}{{Hartigan} et~al.}{2005}]{2005AJ....130.2197H}
{Hartigan} P.,  {Heathcote} S.,  {Morse} J.~A.,  {Reipurth} B.,   {Bally} J.,
  2005, \mn@doi [\aj] {10.1086/491673}, \href
  {https://ui.adsabs.harvard.edu/abs/2005AJ....130.2197H} {130, 2197}

\bibitem[\protect\citeauthoryear{{Haworth} \& {Clarke}}{{Haworth} \&
  {Clarke}}{2019}]{2019MNRAS.485.3895H}
{Haworth} T.~J.,  {Clarke} C.~J.,  2019, \mn@doi [\mnras]
  {10.1093/mnras/stz706}, \href
  {https://ui.adsabs.harvard.edu/abs/2019MNRAS.485.3895H} {485, 3895}

\bibitem[\protect\citeauthoryear{{Haworth} \& {Owen}}{{Haworth} \&
  {Owen}}{2020}]{2020MNRAS.492.5030H}
{Haworth} T.~J.,  {Owen} J.~E.,  2020, \mn@doi [\mnras]
  {10.1093/mnras/staa151}, \href
  {https://ui.adsabs.harvard.edu/abs/2020MNRAS.492.5030H} {492, 5030}

\bibitem[\protect\citeauthoryear{{Haworth}, {Clarke}, {Rahman}, {Winter}  \&
  {Facchini}}{{Haworth} et~al.}{2018}]{2018MNRAS.481..452H}
{Haworth} T.~J.,  {Clarke} C.~J.,  {Rahman} W.,  {Winter} A.~J.,   {Facchini}
  S.,  2018, \mn@doi [\mnras] {10.1093/mnras/sty2323}, \href
  {https://ui.adsabs.harvard.edu/abs/2018MNRAS.481..452H} {481, 452}

\bibitem[\protect\citeauthoryear{{Haworth}, {Kim}, {Winter}, {Hines}, {Clarke},
  {Sellek}, {Ballabio}  \& {Stapelfeldt}}{{Haworth}
  et~al.}{2021}]{2021MNRAS.501.3502H}
{Haworth} T.~J.,  {Kim} J.~S.,  {Winter} A.~J.,  {Hines} D.~C.,  {Clarke}
  C.~J.,  {Sellek} A.~D.,  {Ballabio} G.,   {Stapelfeldt} K.~R.,  2021, \mn@doi
  [\mnras] {10.1093/mnras/staa3918}, \href
  {https://ui.adsabs.harvard.edu/abs/2021MNRAS.501.3502H} {501, 3502}

\bibitem[\protect\citeauthoryear{{Haworth} et~al.,}{{Haworth}
  et~al.}{2022}]{2022MNRAS.512.2594H}
{Haworth} T.~J.,  et~al., 2022, \mn@doi [\mnras] {10.1093/mnras/stac656}, \href
  {https://ui.adsabs.harvard.edu/abs/2022MNRAS.512.2594H} {512, 2594}

\bibitem[\protect\citeauthoryear{{Henney} \& {O'Dell}}{{Henney} \&
  {O'Dell}}{1999}]{1999AJ....118.2350H}
{Henney} W.~J.,  {O'Dell} C.~R.,  1999, \mn@doi [\aj] {10.1086/301087}, \href
  {https://ui.adsabs.harvard.edu/abs/1999AJ....118.2350H} {118, 2350}

\bibitem[\protect\citeauthoryear{{Henney}, {Arthur}  \&
  {Garc{\'\i}a-D{\'\i}az}}{{Henney} et~al.}{2005}]{2005ApJ...627..813H}
{Henney} W.~J.,  {Arthur} S.~J.,   {Garc{\'\i}a-D{\'\i}az} M.~T.,  2005,
  \mn@doi [\apj] {10.1086/430593}, \href
  {https://ui.adsabs.harvard.edu/abs/2005ApJ...627..813H} {627, 813}

\bibitem[\protect\citeauthoryear{{Henroteau} \& {Henderson}}{{Henroteau} \&
  {Henderson}}{1920}]{1920PDO.....5....1H}
{Henroteau} F.,  {Henderson} J.~P.,  1920, Publications of the Dominion
  Observatory Ottawa, \href
  {https://ui.adsabs.harvard.edu/abs/1920PDO.....5....1H} {5, 1}

\bibitem[\protect\citeauthoryear{{Huang} et~al.,}{{Huang}
  et~al.}{2022}]{2022ApJ...930..171H}
{Huang} J.,  et~al., 2022, \mn@doi [\apj] {10.3847/1538-4357/ac63ba}, \href
  {https://ui.adsabs.harvard.edu/abs/2022ApJ...930..171H} {930, 171}

\bibitem[\protect\citeauthoryear{{Jansen}, {Spaans}, {Hogerheijde}  \& {van
  Dishoeck}}{{Jansen} et~al.}{1995}]{1995A&A...303..541J}
{Jansen} D.~J.,  {Spaans} M.,  {Hogerheijde} M.~R.,   {van Dishoeck} E.~F.,
  1995, \aap, \href {https://ui.adsabs.harvard.edu/abs/1995A&A...303..541J}
  {303, 541}

\bibitem[\protect\citeauthoryear{{Jeffries}}{{Jeffries}}{2007}]{2007MNRAS.376.1109J}
{Jeffries} R.~D.,  2007, \mn@doi [\mnras] {10.1111/j.1365-2966.2007.11471.x},
  \href {https://ui.adsabs.harvard.edu/abs/2007MNRAS.376.1109J} {376, 1109}

\bibitem[\protect\citeauthoryear{{Johnstone}, {Hollenbach}  \&
  {Bally}}{{Johnstone} et~al.}{1998}]{1998ApJ...499..758J}
{Johnstone} D.,  {Hollenbach} D.,   {Bally} J.,  1998, \mn@doi [\apj]
  {10.1086/305658}, \href
  {https://ui.adsabs.harvard.edu/abs/1998ApJ...499..758J} {499, 758}

\bibitem[\protect\citeauthoryear{{Keppler} et~al.,}{{Keppler}
  et~al.}{2018}]{2018A&A...617A..44K}
{Keppler} M.,  et~al., 2018, \mn@doi [\aap] {10.1051/0004-6361/201832957},
  \href {https://ui.adsabs.harvard.edu/abs/2018A&A...617A..44K} {617, A44}

\bibitem[\protect\citeauthoryear{{Kim}, {Clarke}, {Fang}  \& {Facchini}}{{Kim}
  et~al.}{2016}]{2016ApJ...826L..15K}
{Kim} J.~S.,  {Clarke} C.~J.,  {Fang} M.,   {Facchini} S.,  2016, \mn@doi
  [\apjl] {10.3847/2041-8205/826/1/L15}, \href
  {https://ui.adsabs.harvard.edu/abs/2016ApJ...826L..15K} {826, L15}

\bibitem[\protect\citeauthoryear{{Kirwan} et~al.,}{{Kirwan}
  et~al.}{2023}]{2023A&A...673A.166K}
{Kirwan} A.,  et~al., 2023, \mn@doi [\aap] {10.1051/0004-6361/202245428}, \href
  {https://ui.adsabs.harvard.edu/abs/2023A&A...673A.166K} {673, A166}

\bibitem[\protect\citeauthoryear{{Kounkel} et~al.,}{{Kounkel}
  et~al.}{2017}]{2017ApJ...834..142K}
{Kounkel} M.,  et~al., 2017, \mn@doi [\apj] {10.3847/1538-4357/834/2/142},
  \href {https://ui.adsabs.harvard.edu/abs/2017ApJ...834..142K} {834, 142}

\bibitem[\protect\citeauthoryear{{Kounkel} et~al.,}{{Kounkel}
  et~al.}{2018}]{2018AJ....156...84K}
{Kounkel} M.,  et~al., 2018, \mn@doi [\aj] {10.3847/1538-3881/aad1f1}, \href
  {https://ui.adsabs.harvard.edu/abs/2018AJ....156...84K} {156, 84}

\bibitem[\protect\citeauthoryear{{Kratter} \& {Lodato}}{{Kratter} \&
  {Lodato}}{2016}]{2016ARA&A..54..271K}
{Kratter} K.,  {Lodato} G.,  2016, \mn@doi [\araa]
  {10.1146/annurev-astro-081915-023307}, \href
  {https://ui.adsabs.harvard.edu/abs/2016ARA&A..54..271K} {54, 271}

\bibitem[\protect\citeauthoryear{{Kurucz}}{{Kurucz}}{1979}]{1979ApJS...40....1K}
{Kurucz} R.~L.,  1979, \mn@doi [\apjs] {10.1086/190589}, \href
  {https://ui.adsabs.harvard.edu/abs/1979ApJS...40....1K} {40, 1}

\bibitem[\protect\citeauthoryear{{Lada} \& {Lada}}{{Lada} \&
  {Lada}}{2003}]{2003ARA&A..41...57L}
{Lada} C.~J.,  {Lada} E.~A.,  2003, \mn@doi [\araa]
  {10.1146/annurev.astro.41.011802.094844}, \href
  {https://ui.adsabs.harvard.edu/abs/2003ARA&A..41...57L} {41, 57}

\bibitem[\protect\citeauthoryear{{Luridiana}, {Morisset}  \&
  {Shaw}}{{Luridiana} et~al.}{2015}]{2015A&A...573A..42L}
{Luridiana} V.,  {Morisset} C.,   {Shaw} R.~A.,  2015, \mn@doi [\aap]
  {10.1051/0004-6361/201323152}, \href
  {https://ui.adsabs.harvard.edu/abs/2015A&A...573A..42L} {573, A42}

\bibitem[\protect\citeauthoryear{{Mann} et~al.,}{{Mann}
  et~al.}{2014}]{2014ApJ...784...82M}
{Mann} R.~K.,  et~al., 2014, \mn@doi [\apj] {10.1088/0004-637X/784/1/82}, \href
  {https://ui.adsabs.harvard.edu/abs/2014ApJ...784...82M} {784, 82}

\bibitem[\protect\citeauthoryear{{McLeod}, {Weilbacher}, {Ginsburg}, {Dale},
  {Ramsay}  \& {Testi}}{{McLeod} et~al.}{2016}]{2016MNRAS.455.4057M}
{McLeod} A.~F.,  {Weilbacher} P.~M.,  {Ginsburg} A.,  {Dale} J.~E.,  {Ramsay}
  S.,   {Testi} L.,  2016, \mn@doi [\mnras] {10.1093/mnras/stv2617}, \href
  {https://ui.adsabs.harvard.edu/abs/2016MNRAS.455.4057M} {455, 4057}

\bibitem[\protect\citeauthoryear{{Mesa-Delgado}, {N{\'u}{\~n}ez-D{\'\i}az},
  {Esteban}, {L{\'o}pez-Mart{\'\i}n}  \& {Garc{\'\i}a-Rojas}}{{Mesa-Delgado}
  et~al.}{2011}]{2011MNRAS.417..420M}
{Mesa-Delgado} A.,  {N{\'u}{\~n}ez-D{\'\i}az} M.,  {Esteban} C.,
  {L{\'o}pez-Mart{\'\i}n} L.,   {Garc{\'\i}a-Rojas} J.,  2011, \mn@doi [\mnras]
  {10.1111/j.1365-2966.2011.19278.x}, \href
  {https://ui.adsabs.harvard.edu/abs/2011MNRAS.417..420M} {417, 420}

\bibitem[\protect\citeauthoryear{{Miotello}, {Robberto}, {Potenza}  \&
  {Ricci}}{{Miotello} et~al.}{2012}]{2012ApJ...757...78M}
{Miotello} A.,  {Robberto} M.,  {Potenza} M. A.~C.,   {Ricci} L.,  2012,
  \mn@doi [\apj] {10.1088/0004-637X/757/1/78}, \href
  {https://ui.adsabs.harvard.edu/abs/2012ApJ...757...78M} {757, 78}

\bibitem[\protect\citeauthoryear{{Miotello}, {Rosotti}, {Ansdell}, {Facchini},
  {Manara}, {Williams}  \& {Bruderer}}{{Miotello}
  et~al.}{2021}]{2021A&A...651A..48M}
{Miotello} A.,  {Rosotti} G.,  {Ansdell} M.,  {Facchini} S.,  {Manara} C.~F.,
  {Williams} J.~P.,   {Bruderer} S.,  2021, \mn@doi [\aap]
  {10.1051/0004-6361/202140550}, \href
  {https://ui.adsabs.harvard.edu/abs/2021A&A...651A..48M} {651, A48}

\bibitem[\protect\citeauthoryear{{Nisini}, {Antoniucci}, {Alcal{\'a}},
  {Giannini}, {Manara}, {Natta}, {Fedele}  \& {Biazzo}}{{Nisini}
  et~al.}{2018}]{2018A&A...609A..87N}
{Nisini} B.,  {Antoniucci} S.,  {Alcal{\'a}} J.~M.,  {Giannini} T.,  {Manara}
  C.~F.,  {Natta} A.,  {Fedele} D.,   {Biazzo} K.,  2018, \mn@doi [\aap]
  {10.1051/0004-6361/201730834}, \href
  {https://ui.adsabs.harvard.edu/abs/2018A&A...609A..87N} {609, A87}

\bibitem[\protect\citeauthoryear{{O'Dell}}{{O'Dell}}{2018}]{2018MNRAS.478.1017O}
{O'Dell} C.~R.,  2018, \mn@doi [\mnras] {10.1093/mnras/sty960}, \href
  {https://ui.adsabs.harvard.edu/abs/2018MNRAS.478.1017O} {478, 1017}

\bibitem[\protect\citeauthoryear{{O'Dell} \& {Harris}}{{O'Dell} \&
  {Harris}}{2010}]{2010AJ....140..985O}
{O'Dell} C.~R.,  {Harris} J.~A.,  2010, \mn@doi [\aj]
  {10.1088/0004-6256/140/4/985}, \href
  {https://ui.adsabs.harvard.edu/abs/2010AJ....140..985O} {140, 985}

\bibitem[\protect\citeauthoryear{{O'Dell} \& {Wen}}{{O'Dell} \&
  {Wen}}{1994}]{1994ApJ...436..194O}
{O'Dell} C.~R.,  {Wen} Z.,  1994, \mn@doi [\apj] {10.1086/174892}, \href
  {https://ui.adsabs.harvard.edu/abs/1994ApJ...436..194O} {436, 194}

\bibitem[\protect\citeauthoryear{{O'Dell}, {Wen}  \& {Hu}}{{O'Dell}
  et~al.}{1993}]{1993ApJ...410..696O}
{O'Dell} C.~R.,  {Wen} Z.,   {Hu} X.,  1993, \mn@doi [\apj] {10.1086/172786},
  \href {https://ui.adsabs.harvard.edu/abs/1993ApJ...410..696O} {410, 696}

\bibitem[\protect\citeauthoryear{{O'Dell}, {Henney}, {Abel}, {Ferland}  \&
  {Arthur}}{{O'Dell} et~al.}{2009}]{2009AJ....137..367O}
{O'Dell} C.~R.,  {Henney} W.~J.,  {Abel} N.~P.,  {Ferland} G.~J.,   {Arthur}
  S.~J.,  2009, \mn@doi [\aj] {10.1088/0004-6256/137/1/367}, \href
  {https://ui.adsabs.harvard.edu/abs/2009AJ....137..367O} {137, 367}

\bibitem[\protect\citeauthoryear{{O'Dell}, {Ferland}  \& {Peimbert}}{{O'Dell}
  et~al.}{2017a}]{2017MNRAS.464.4835O}
{O'Dell} C.~R.,  {Ferland} G.~J.,   {Peimbert} M.,  2017a, \mn@doi [\mnras]
  {10.1093/mnras/stw2713}, \href
  {https://ui.adsabs.harvard.edu/abs/2017MNRAS.464.4835O} {464, 4835}

\bibitem[\protect\citeauthoryear{{O'Dell}, {Kollatschny}  \&
  {Ferland}}{{O'Dell} et~al.}{2017b}]{2017ApJ...837..151O}
{O'Dell} C.~R.,  {Kollatschny} W.,   {Ferland} G.~J.,  2017b, \mn@doi [\apj]
  {10.3847/1538-4357/aa6198}, \href
  {https://ui.adsabs.harvard.edu/abs/2017ApJ...837..151O} {837, 151}

\bibitem[\protect\citeauthoryear{{O'Dell}, {Abel}  \& {Ferland}}{{O'Dell}
  et~al.}{2020}]{2020ApJ...891...46O}
{O'Dell} C.~R.,  {Abel} N.~P.,   {Ferland} G.~J.,  2020, \mn@doi [\apj]
  {10.3847/1538-4357/ab723d}, \href
  {https://ui.adsabs.harvard.edu/abs/2020ApJ...891...46O} {891, 46}

\bibitem[\protect\citeauthoryear{{{\"O}berg} et~al.,}{{{\"O}berg}
  et~al.}{2021}]{2021ApJS..257....1O}
{{\"O}berg} K.~I.,  et~al., 2021, \mn@doi [\apjs] {10.3847/1538-4365/ac1432},
  \href {https://ui.adsabs.harvard.edu/abs/2021ApJS..257....1O} {257, 1}

\bibitem[\protect\citeauthoryear{{O'dell} \& {Wong}}{{O'dell} \&
  {Wong}}{1996}]{1996AJ....111..846O}
{O'dell} C.~R.,  {Wong} K.,  1996, \mn@doi [\aj] {10.1086/117832}, \href
  {https://ui.adsabs.harvard.edu/abs/1996AJ....111..846O} {111, 846}

\bibitem[\protect\citeauthoryear{{Osterbrock} \& {Ferland}}{{Osterbrock} \&
  {Ferland}}{2006}]{2006agna.book.....O}
{Osterbrock} D.~E.,  {Ferland} G.~J.,  2006, {Astrophysics of gaseous nebulae
  and active galactic nuclei}

\bibitem[\protect\citeauthoryear{{Osterbrock}, {Tran}  \&
  {Veilleux}}{{Osterbrock} et~al.}{1992}]{1992ApJ...389..305O}
{Osterbrock} D.~E.,  {Tran} H.~D.,   {Veilleux} S.,  1992, \mn@doi [\apj]
  {10.1086/171206}, \href
  {https://ui.adsabs.harvard.edu/abs/1992ApJ...389..305O} {389, 305}

\bibitem[\protect\citeauthoryear{{Pesenti}, {Dougados}, {Cabrit}, {O'Brien},
  {Garcia}  \& {Ferreira}}{{Pesenti} et~al.}{2003}]{2003A&A...410..155P}
{Pesenti} N.,  {Dougados} C.,  {Cabrit} S.,  {O'Brien} D.,  {Garcia} P.,
  {Ferreira} J.,  2003, \mn@doi [\aap] {10.1051/0004-6361:20031131}, \href
  {https://ui.adsabs.harvard.edu/abs/2003A&A...410..155P} {410, 155}

\bibitem[\protect\citeauthoryear{{Pfalzner}}{{Pfalzner}}{2013}]{2013A&A...549A..82P}
{Pfalzner} S.,  2013, \mn@doi [\aap] {10.1051/0004-6361/201218792}, \href
  {https://ui.adsabs.harvard.edu/abs/2013A&A...549A..82P} {549, A82}

\bibitem[\protect\citeauthoryear{{Pfalzner}, {Umbreit}  \&
  {Henning}}{{Pfalzner} et~al.}{2005}]{2005ApJ...629..526P}
{Pfalzner} S.,  {Umbreit} S.,   {Henning} T.,  2005, \mn@doi [\apj]
  {10.1086/431350}, \href
  {https://ui.adsabs.harvard.edu/abs/2005ApJ...629..526P} {629, 526}

\bibitem[\protect\citeauthoryear{{Pfalzner}, {Bhandare}, {Vincke}  \&
  {Lacerda}}{{Pfalzner} et~al.}{2018}]{2018ApJ...863...45P}
{Pfalzner} S.,  {Bhandare} A.,  {Vincke} K.,   {Lacerda} P.,  2018, \mn@doi
  [\apj] {10.3847/1538-4357/aad23c}, \href
  {https://ui.adsabs.harvard.edu/abs/2018ApJ...863...45P} {863, 45}

\bibitem[\protect\citeauthoryear{{Pineda}, {Segura-Cox}, {Caselli},
  {Cunningham}, {Zhao}, {Schmiedeke}, {Maureira}  \& {Neri}}{{Pineda}
  et~al.}{2020}]{2020NatAs...4.1158P}
{Pineda} J.~E.,  {Segura-Cox} D.,  {Caselli} P.,  {Cunningham} N.,  {Zhao} B.,
  {Schmiedeke} A.,  {Maureira} M.~J.,   {Neri} R.,  2020, \mn@doi [Nature
  Astronomy] {10.1038/s41550-020-1150-z}, \href
  {https://ui.adsabs.harvard.edu/abs/2020NatAs...4.1158P} {4, 1158}

\bibitem[\protect\citeauthoryear{{Pinte} et~al.,}{{Pinte}
  et~al.}{2018}]{2018ApJ...860L..13P}
{Pinte} C.,  et~al., 2018, \mn@doi [\apjl] {10.3847/2041-8213/aac6dc}, \href
  {https://ui.adsabs.harvard.edu/abs/2018ApJ...860L..13P} {860, L13}

\bibitem[\protect\citeauthoryear{{Pinte} et~al.,}{{Pinte}
  et~al.}{2020}]{2020ApJ...890L...9P}
{Pinte} C.,  et~al., 2020, \mn@doi [\apjl] {10.3847/2041-8213/ab6dda}, \href
  {https://ui.adsabs.harvard.edu/abs/2020ApJ...890L...9P} {890, L9}

\bibitem[\protect\citeauthoryear{{Pinte}, {Teague}, {Flaherty}, {Hall},
  {Facchini}  \& {Casassus}}{{Pinte} et~al.}{2022}]{2022arXiv220309528P}
{Pinte} C.,  {Teague} R.,  {Flaherty} K.,  {Hall} C.,  {Facchini} S.,
  {Casassus} S.,  2022, \mn@doi [arXiv e-prints] {10.48550/arXiv.2203.09528},
  \href {https://ui.adsabs.harvard.edu/abs/2022arXiv220309528P} {p.
  arXiv:2203.09528}

\bibitem[\protect\citeauthoryear{{Portegies Zwart}, {McMillan}  \&
  {Gieles}}{{Portegies Zwart} et~al.}{2010}]{2010ARA&A..48..431P}
{Portegies Zwart} S.~F.,  {McMillan} S. L.~W.,   {Gieles} M.,  2010, \mn@doi
  [\araa] {10.1146/annurev-astro-081309-130834}, \href
  {https://ui.adsabs.harvard.edu/abs/2010ARA&A..48..431P} {48, 431}

\bibitem[\protect\citeauthoryear{{Qiao}, {Haworth}, {Sellek}  \& {Ali}}{{Qiao}
  et~al.}{2022}]{2022MNRAS.512.3788Q}
{Qiao} L.,  {Haworth} T.~J.,  {Sellek} A.~D.,   {Ali} A.~A.,  2022, \mn@doi
  [\mnras] {10.1093/mnras/stac684}, \href
  {https://ui.adsabs.harvard.edu/abs/2022MNRAS.512.3788Q} {512, 3788}

\bibitem[\protect\citeauthoryear{{Qiao}, {Coleman}  \& {Haworth}}{{Qiao}
  et~al.}{2023}]{2023MNRAS.522.1939Q}
{Qiao} L.,  {Coleman} G. A.~L.,   {Haworth} T.~J.,  2023, \mn@doi [\mnras]
  {10.1093/mnras/stad944}, \href
  {https://ui.adsabs.harvard.edu/abs/2023MNRAS.522.1939Q} {522, 1939}

\bibitem[\protect\citeauthoryear{{Reiter} \& {Parker}}{{Reiter} \&
  {Parker}}{2022}]{2022EPJP..137.1071R}
{Reiter} M.,  {Parker} R.~J.,  2022, \mn@doi [European Physical Journal Plus]
  {10.1140/epjp/s13360-022-03265-7}, \href
  {https://ui.adsabs.harvard.edu/abs/2022EPJP..137.1071R} {137, 1071}

\bibitem[\protect\citeauthoryear{{Reiter} \& {Smith}}{{Reiter} \&
  {Smith}}{2013}]{2013MNRAS.433.2226R}
{Reiter} M.,  {Smith} N.,  2013, \mn@doi [\mnras] {10.1093/mnras/stt889}, \href
  {https://ui.adsabs.harvard.edu/abs/2013MNRAS.433.2226R} {433, 2226}

\bibitem[\protect\citeauthoryear{{Reiter}, {Smith}, {Kiminki}  \&
  {Bally}}{{Reiter} et~al.}{2015}]{2015MNRAS.450..564R}
{Reiter} M.,  {Smith} N.,  {Kiminki} M.~M.,   {Bally} J.,  2015, \mn@doi
  [\mnras] {10.1093/mnras/stv634}, \href
  {https://ui.adsabs.harvard.edu/abs/2015MNRAS.450..564R} {450, 564}

\bibitem[\protect\citeauthoryear{{Reiter}, {Smith}  \& {Bally}}{{Reiter}
  et~al.}{2016}]{2016MNRAS.463.4344R}
{Reiter} M.,  {Smith} N.,   {Bally} J.,  2016, \mn@doi [\mnras]
  {10.1093/mnras/stw2296}, \href
  {https://ui.adsabs.harvard.edu/abs/2016MNRAS.463.4344R} {463, 4344}

\bibitem[\protect\citeauthoryear{{Reiter}, {Kiminki}, {Smith}  \&
  {Bally}}{{Reiter} et~al.}{2017}]{2017MNRAS.470.4671R}
{Reiter} M.,  {Kiminki} M.~M.,  {Smith} N.,   {Bally} J.,  2017, \mn@doi
  [\mnras] {10.1093/mnras/stx1489}, \href
  {https://ui.adsabs.harvard.edu/abs/2017MNRAS.470.4671R} {470, 4671}

\bibitem[\protect\citeauthoryear{{Reiter}, {McLeod}, {Klaassen}, {Guzm{\'a}n},
  {Dale}, {Mottram}  \& {Garay}}{{Reiter} et~al.}{2019}]{2019MNRAS.490.2056R}
{Reiter} M.,  {McLeod} A.~F.,  {Klaassen} P.~D.,  {Guzm{\'a}n} A.~E.,  {Dale}
  J.~E.,  {Mottram} J.~C.,   {Garay} G.,  2019, \mn@doi [\mnras]
  {10.1093/mnras/stz2752}, \href
  {https://ui.adsabs.harvard.edu/abs/2019MNRAS.490.2056R} {490, 2056}

\bibitem[\protect\citeauthoryear{{Reiter}, {Morse}, {Smith}, {Haworth}, {Kuhn}
  \& {Klaassen}}{{Reiter} et~al.}{2022}]{2022MNRAS.517.5382R}
{Reiter} M.,  {Morse} J.~A.,  {Smith} N.,  {Haworth} T.~J.,  {Kuhn} M.~A.,
  {Klaassen} P.~D.,  2022, \mn@doi [\mnras] {10.1093/mnras/stac2820}, \href
  {https://ui.adsabs.harvard.edu/abs/2022MNRAS.517.5382R} {517, 5382}

\bibitem[\protect\citeauthoryear{{Ricci}, {Robberto}  \& {Soderblom}}{{Ricci}
  et~al.}{2008}]{2008AJ....136.2136R}
{Ricci} L.,  {Robberto} M.,   {Soderblom} D.~R.,  2008, \mn@doi [\aj]
  {10.1088/0004-6256/136/5/2136}, \href
  {https://ui.adsabs.harvard.edu/abs/2008AJ....136.2136R} {136, 2136}

\bibitem[\protect\citeauthoryear{{Richling} \& {Yorke}}{{Richling} \&
  {Yorke}}{2000}]{2000ApJ...539..258R}
{Richling} S.,  {Yorke} H.~W.,  2000, \mn@doi [\apj] {10.1086/309198}, \href
  {https://ui.adsabs.harvard.edu/abs/2000ApJ...539..258R} {539, 258}

\bibitem[\protect\citeauthoryear{{Rubin}, {Simpson}, {O'Dell}, {McNabb},
  {Colgan}, {Zhuge}, {Ferland}  \& {Hidalgo}}{{Rubin}
  et~al.}{2011}]{2011MNRAS.410.1320R}
{Rubin} R.~H.,  {Simpson} J.~P.,  {O'Dell} C.~R.,  {McNabb} I.~A.,  {Colgan} S.
  W.~J.,  {Zhuge} S.~Y.,  {Ferland} G.~J.,   {Hidalgo} S.~A.,  2011, \mn@doi
  [\mnras] {10.1111/j.1365-2966.2010.17522.x}, \href
  {https://ui.adsabs.harvard.edu/abs/2011MNRAS.410.1320R} {410, 1320}

\bibitem[\protect\citeauthoryear{{Schulz}, {Testa}, {Huenemoerder}, {Ishibashi}
   \& {Canizares}}{{Schulz} et~al.}{2006}]{2006ApJ...653..636S}
{Schulz} N.~S.,  {Testa} P.,  {Huenemoerder} D.~P.,  {Ishibashi} K.,
  {Canizares} C.~R.,  2006, \mn@doi [\apj] {10.1086/508625}, \href
  {https://ui.adsabs.harvard.edu/abs/2006ApJ...653..636S} {653, 636}

\bibitem[\protect\citeauthoryear{{Sicilia-Aguilar} et~al.,}{{Sicilia-Aguilar}
  et~al.}{2005}]{2005AJ....129..363S}
{Sicilia-Aguilar} A.,  et~al., 2005, \mn@doi [\aj] {10.1086/426327}, \href
  {https://ui.adsabs.harvard.edu/abs/2005AJ....129..363S} {129, 363}

\bibitem[\protect\citeauthoryear{{Storey} \& {Hummer}}{{Storey} \&
  {Hummer}}{1995}]{1995MNRAS.272...41S}
{Storey} P.~J.,  {Hummer} D.~G.,  1995, \mn@doi [\mnras]
  {10.1093/mnras/272.1.41}, \href
  {https://ui.adsabs.harvard.edu/abs/1995MNRAS.272...41S} {272, 41}

\bibitem[\protect\citeauthoryear{{St{\"o}rzer} \& {Hollenbach}}{{St{\"o}rzer}
  \& {Hollenbach}}{1998}]{1998ApJ...502L..71S}
{St{\"o}rzer} H.,  {Hollenbach} D.,  1998, \mn@doi [\apjl] {10.1086/311487},
  \href {https://ui.adsabs.harvard.edu/abs/1998ApJ...502L..71S} {502, L71}

\bibitem[\protect\citeauthoryear{{Tang}, {Guilloteau}, {Pi{\'e}tu}, {Dutrey},
  {Ohashi}  \& {Ho}}{{Tang} et~al.}{2012}]{2012A&A...547A..84T}
{Tang} Y.~W.,  {Guilloteau} S.,  {Pi{\'e}tu} V.,  {Dutrey} A.,  {Ohashi} N.,
  {Ho} P.~T.~P.,  2012, \mn@doi [\aap] {10.1051/0004-6361/201219414}, \href
  {https://ui.adsabs.harvard.edu/abs/2012A&A...547A..84T} {547, A84}

\bibitem[\protect\citeauthoryear{{Tauber}, {Tielens}, {Meixner}  \&
  {Goldsmith}}{{Tauber} et~al.}{1994}]{1994ApJ...422..136T}
{Tauber} J.~A.,  {Tielens} A.~G.~G.~M.,  {Meixner} M.,   {Goldsmith} P.~F.,
  1994, \mn@doi [\apj] {10.1086/173711}, \href
  {https://ui.adsabs.harvard.edu/abs/1994ApJ...422..136T} {422, 136}

\bibitem[\protect\citeauthoryear{{Teague}, {Bae}, {Bergin}, {Birnstiel}  \&
  {Foreman-Mackey}}{{Teague} et~al.}{2018}]{2018ApJ...860L..12T}
{Teague} R.,  {Bae} J.,  {Bergin} E.~A.,  {Birnstiel} T.,   {Foreman-Mackey}
  D.,  2018, \mn@doi [\apjl] {10.3847/2041-8213/aac6d7}, \href
  {https://ui.adsabs.harvard.edu/abs/2018ApJ...860L..12T} {860, L12}

\bibitem[\protect\citeauthoryear{{Valdivia-Mena} et~al.,}{{Valdivia-Mena}
  et~al.}{2022}]{2022A&A...667A..12V}
{Valdivia-Mena} M.~T.,  et~al., 2022, \mn@doi [\aap]
  {10.1051/0004-6361/202243310}, \href
  {https://ui.adsabs.harvard.edu/abs/2022A&A...667A..12V} {667, A12}

\bibitem[\protect\citeauthoryear{{Walch}, {Whitworth}, {Bisbas}, {W{\"u}nsch}
  \& {Hubber}}{{Walch} et~al.}{2012}]{2012MNRAS.427..625W}
{Walch} S.~K.,  {Whitworth} A.~P.,  {Bisbas} T.,  {W{\"u}nsch} R.,   {Hubber}
  D.,  2012, \mn@doi [\mnras] {10.1111/j.1365-2966.2012.21767.x}, \href
  {https://ui.adsabs.harvard.edu/abs/2012MNRAS.427..625W} {427, 625}

\bibitem[\protect\citeauthoryear{{Walmsley}, {Natta}, {Oliva}  \&
  {Testi}}{{Walmsley} et~al.}{2000}]{2000A&A...364..301W}
{Walmsley} C.~M.,  {Natta} A.,  {Oliva} E.,   {Testi} L.,  2000, \aap, \href
  {https://ui.adsabs.harvard.edu/abs/2000A&A...364..301W} {364, 301}

\bibitem[\protect\citeauthoryear{{Weilbacher} et~al.,}{{Weilbacher}
  et~al.}{2015}]{2015A&A...582A.114W}
{Weilbacher} P.~M.,  et~al., 2015, \mn@doi [\aap]
  {10.1051/0004-6361/201526529}, \href
  {https://ui.adsabs.harvard.edu/abs/2015A&A...582A.114W} {582, A114}

\bibitem[\protect\citeauthoryear{{Wen} \& {O'Dell}}{{Wen} \&
  {O'Dell}}{1995}]{1995ApJ...438..784W}
{Wen} Z.,  {O'Dell} C.~R.,  1995, \mn@doi [\apj] {10.1086/175123}, \href
  {https://ui.adsabs.harvard.edu/abs/1995ApJ...438..784W} {438, 784}

\bibitem[\protect\citeauthoryear{{Wilhelm}, {Portegies Zwart},
  {Cournoyer-Cloutier}, {Lewis}, {Polak}, {Tran}  \& {Mac Low}}{{Wilhelm}
  et~al.}{2023}]{2023MNRAS.520.5331W}
{Wilhelm} M. J.~C.,  {Portegies Zwart} S.,  {Cournoyer-Cloutier} C.,  {Lewis}
  S.~C.,  {Polak} B.,  {Tran} A.,   {Mac Low} M.-M.,  2023, \mn@doi [\mnras]
  {10.1093/mnras/stad445}, \href
  {https://ui.adsabs.harvard.edu/abs/2023MNRAS.520.5331W} {520, 5331}

\bibitem[\protect\citeauthoryear{{Winter} \& {Haworth}}{{Winter} \&
  {Haworth}}{2022}]{2022EPJP..137.1132W}
{Winter} A.~J.,  {Haworth} T.~J.,  2022, \mn@doi [European Physical Journal
  Plus] {10.1140/epjp/s13360-022-03314-1}, \href
  {https://ui.adsabs.harvard.edu/abs/2022EPJP..137.1132W} {137, 1132}

\bibitem[\protect\citeauthoryear{{Winter}, {Clarke}, {Rosotti}, {Hacar}  \&
  {Alexander}}{{Winter} et~al.}{2019}]{2019MNRAS.490.5478W}
{Winter} A.~J.,  {Clarke} C.~J.,  {Rosotti} G.~P.,  {Hacar} A.,   {Alexander}
  R.,  2019, \mn@doi [\mnras] {10.1093/mnras/stz2545}, \href
  {https://ui.adsabs.harvard.edu/abs/2019MNRAS.490.5478W} {490, 5478}

\bibitem[\protect\citeauthoryear{{Yen}, {Gu}, {Hirano}, {Koch}, {Lee}, {Liu}
  \& {Takakuwa}}{{Yen} et~al.}{2019}]{2019ApJ...880...69Y}
{Yen} H.-W.,  {Gu} P.-G.,  {Hirano} N.,  {Koch} P.~M.,  {Lee} C.-F.,  {Liu}
  H.~B.,   {Takakuwa} S.,  2019, \mn@doi [\apj] {10.3847/1538-4357/ab29f8},
  \href {https://ui.adsabs.harvard.edu/abs/2019ApJ...880...69Y} {880, 69}

\bibitem[\protect\citeauthoryear{{Zeidler}}{{Zeidler}}{2019}]{2019zndo...3433996Z}
{Zeidler} P.,  2019, {MUSEpack}, Zenodo, \mn@doi{10.5281/zenodo.3433996}

\bibitem[\protect\citeauthoryear{{Zeidler}, {Nota}, {Sabbi}, {Luljak},
  {McLeod}, {Grebel}, {Pasquali}  \& {Tosi}}{{Zeidler}
  et~al.}{2019}]{2019AJ....158..201Z}
{Zeidler} P.,  {Nota} A.,  {Sabbi} E.,  {Luljak} P.,  {McLeod} A.~F.,  {Grebel}
  E.~K.,  {Pasquali} A.,   {Tosi} M.,  2019, \mn@doi [\aj]
  {10.3847/1538-3881/ab44bb}, \href
  {https://ui.adsabs.harvard.edu/abs/2019AJ....158..201Z} {158, 201}

\bibitem[\protect\citeauthoryear{{Zhang} et~al.,}{{Zhang}
  et~al.}{2018}]{2018ApJ...869L..47Z}
{Zhang} S.,  et~al., 2018, \mn@doi [\apjl] {10.3847/2041-8213/aaf744}, \href
  {https://ui.adsabs.harvard.edu/abs/2018ApJ...869L..47Z} {869, L47}

\bibitem[\protect\citeauthoryear{{Zhang}, {Kalscheur}, {Long}, {Zhang}, {Long},
  {Bergin}, {Zhu}  \& {Trapman}}{{Zhang} et~al.}{2023}]{2023arXiv230503862Z}
{Zhang} S.,  {Kalscheur} M.,  {Long} F.,  {Zhang} K.,  {Long} D.~E.,  {Bergin}
  E.~A.,  {Zhu} Z.,   {Trapman} L.,  2023, \mn@doi [arXiv e-prints]
  {10.48550/arXiv.2305.03862}, \href
  {https://ui.adsabs.harvard.edu/abs/2023arXiv230503862Z} {p. arXiv:2305.03862}

\bibitem[\protect\citeauthoryear{{Zucker} et~al.,}{{Zucker}
  et~al.}{2022}]{2022Natur.601..334Z}
{Zucker} C.,  et~al., 2022, \mn@doi [\nat] {10.1038/s41586-021-04286-5}, \href
  {https://ui.adsabs.harvard.edu/abs/2022Natur.601..334Z} {601, 334}

\makeatother
\end{thebibliography}

%%%%%%%%%%%%%%%%%%%%%%%%%%%%%%%%%%%%%%%%%%%%%%%%%%

% Don't change these lines
\bsp	% typesetting comment
\label{lastpage}
\end{document}